\documentclass[12pt,preprint]{aastex}
\usepackage[numberedappendix]{emulateapj5}
\newcommand{\et}{\mbox{et~al.\ }}
\newcommand{\eg}{\mbox{e.g.,}\ }
\newcommand{\ie}{\mbox{i.e.,}\ }

\newcommand{\kms}{\ifmmode {\rm km\,s}^{-1} \else km\,s$^{-1}$\fi}
\newcommand{\kmsMpc}{\ifmmode {\rm km\,s}^{-1} {\rm Mpc}^{-1} \else km\,s$^{-1}$\,Mpc$^{-1}$\fi}
\newcommand{\ergs}{\ifmmode {\rm ergs\,s}^{-1} \else ergs\,s$^{-1}$\fi}
\newcommand{\lya}{\ifmmode {\rm Ly\,}\alpha \, \else Ly\,$\alpha$\,\fi}
\newcommand{\nv}{N\,{\sc v}} 
\newcommand{\niv}{N\,{\sc iv}]} 
\newcommand{\hb}{H\,$\beta$}

\newcommand{\mgii}{Mg\,{\sc ii}} 
 
\newcommand{\civ}{C\,{\sc iv}} 
\newcommand{\ciii}{C\,{\sc iii}]}
\newcommand{\siiv}{Si\,{\sc iv}}

\newcommand{\heii}{He\,{\sc ii}}
\newcommand{\lam}{$\lambda$}
\newcommand{\mstar}{$M^{\ast}_{\rm BH}$}

\newcommand{\mbh}{$M_{\rm BH}$}
\newcommand{\lbol}{$L_{\rm bol}$}
\newcommand{\lol}{$L_{\rm bol}/L_{\rm Edd}$}
\newcommand{\lstar}{$L^{\ast}_B$}
\newcommand{\lbolstar}{$L^{\ast}_{\rm bol}$}
\newcommand{\leddstar}{$L^{\ast}_{\rm Edd}$}

\newcommand{\Msol}{\mbox{$M_{\odot}$}}

\newcommand{\chandra}{{\it Chandra}}
\newcommand{\rosat}{{\it ROSAT}}
\newcommand{\hst}{{\it HST}}

\newcommand{\lsim}{\stackrel{\scriptscriptstyle <}{\scriptstyle {}_\sim}}
\newcommand{\gsim}{\stackrel{\scriptscriptstyle >}{\scriptstyle {}_\sim}}

\input epsf

\begin{document}
\submitted{Accepted by ApJ, Sept. 14, 2003}
\journalinfo{To appear in The Astrophysical Journal, 600, January 1, 2004}

\title{Early Growth and Efficient Accretion of Massive Black Holes at High 
Redshift}

\shorttitle{Massive Black Holes at High Redshift.}
\shortauthors{Vestergaard}

\author{M.\ Vestergaard}
\affil{Department of Astronomy, The Ohio State University,
	140 West 18th Avenue, \\ Columbus, OH 43210-1173.
	Email: vester@astronomy.ohio-state.edu
}

\begin{abstract}
Black-hole masses of the highest redshift quasars ($4 \lsim z \lsim 6$) 
are estimated using a previously presented scaling relationship, 
derived from reverberation mapping of nearby quasars, and compared to 
quasars at lower redshift.  It is shown that the central black holes in 
luminous $z \gsim 4$ quasars are very massive ($\gsim 10^9$\Msol).
It is argued that the mass estimates of the high-$z$ quasars are 
not subject to larger uncertainties than those for nearby quasars. 
Specifically, the large masses are not overestimates and the lack of 
similarly large black-hole masses in the nearby Universe does not 
rule out their existence at high-$z$.  
However, AGN host galaxies do not typically appear fully formed and/or 
evolved at these early epochs.  This supports scenarios in which black 
holes build up mass very fast in a radiatively inefficient (or obscured) 
phase relative to the stars in their galaxies.  
Additionally, upper envelopes of \mbh{}\,$\approx 10^{10}$\Msol{}
and \lbol{}\,$\approx 10^{48}$ \ergs{} are observed at all redshifts.
\end{abstract}


\section{Introduction \label{introduction}}

High-redshift quasars are of immediate interest to both cosmologists and active
galactic nuclei (AGNs) specialists for several reasons. For example, quasars
can be observed out to greater distances and with more ease than quiescent 
galaxies, owing to their unparalleled energy output over essentially the entire 
electromagnetic spectrum (\eg Elvis \et 1994). 
Since quasars reside in the central regions of massive galaxies (\eg Boroson 
\& Oke 1982; Boroson, Oke, \& Green 1982; Kukula \et 2001), they provide 
a valuable beacon to locate galaxies at early times, which is difficult 
by other means\footnote{However, {\it SIRTF}, {\it JWST}, {\it FIRST/Herschel} and 
ALMA will undoubtedly change this (\eg Neufeld 1999; Omont 2003).} 
(Lanzetta \et 1999). 
Furthermore, quasar activity is intimately connected with the hierarchical 
buildup of galaxies (\eg 
Haehnelt, Natarajan, \& Rees 1998; Cattaneo, Haehnelt, \& Rees 1999; 
Ferrarese \& Merritt 2000; Gebhardt \et 2000a; 
Haehnelt \& Kauffmann 2000; 
Colina \et 2001; Menou, Haiman, \& Narayanan 2001). 
In fact, quasars possibly play a profound role for the formation and evolution of 
most, if not all, galaxies (\eg Kauffmann \& Haehnelt 2000; 
Milosavljevi{\'c} \& Merritt 2001, 2002; Benson \et 2002). 
The quasar phase may even precede the formation of the galaxies themselves
(Rix \et 2001; Kauffmann \& Haehnelt 2000; Omont \et 2001)
and impact their early formation and evolution 
(\eg Silk \& Rees 1998; Haiman, Madau, \& Loeb 1999; 
Theuns, Mo, \& Schaye 2001; Di Matteo \et 2003).

By virtue of being associated with the highest density peaks in the matter 
distribution
(\eg Efstathiou \& Rees 1988; Volonteri, Haardt, \& Madau 2002), quasars 
are an easy probe of these peaks at the earliest epochs and hence of some of the 
first structures formed in the early Universe (\eg Haiman \& Loeb 1998).
In addition, quasars help probe the nature of the medium in which they are
embedded. For example, the spectra of $z \approx 6$ quasars place constraints
on the epoch of re-ionization (Fan \et 2002).
The presence of luminous, massive\footnote{This is typically estimated via 
the Eddington argument but these high black-hole masses are also confirmed
in this work.} quasars at redshift 6, when the Universe was less than 10\% of its 
present age, already place much-needed constraints on models of the earliest 
structure formation (Turner 1991; Haiman \& Loeb 2001; Volonteri \et 2002).

High-redshift quasars are also studied (a) to track the growth and evolution 
of the central black holes (Haiman \& Loeb 2001; Merritt \& Ferrarese 2001a; 
Ferrarese 2002; A. Steed et al. 2003, in preparation), (b) to understand the 
central engine (\eg Peterson, Polidan, \& Pogge 2001; Peterson 2002) 
responsible for their monstrous energy release, and (c) to understand the 
evolution of the quasars and active galaxies in general (\eg Pei 1995; Boyle 
\et 2000; see Hartwick \& Schade 1990 and Osmer 2003 for reviews). 

It is well established that quasars and their activity reign\footnote{Quasars 
were most numerous at the epoch at $z \approx 2.5$, yet they are still rare 
objects (\eg Hartwick \& Schade 1990) relative to the galaxy population (\eg 
Huchra \& Burg 1992).} between redshift 2 and 3
(Osmer 1982; Hartwick \& Schade 1990; Warren, Hewett, \& Osmer 1994; Schmidt, 
Schneider, \& Gunn 1995; Kennefick, Djorgovski, \& Meylan 1996; Fan \et 2001a). 
The decline in comoving space density above redshifts of $\sim 3.5$ is significant not 
only for the optically selected quasars but also for the radio-selected subset 
(Shaver \et 1996; Hook \et 2002) and for X-ray selected sources (Hasinger 2002);
the latter may decline more slowly with increasing redshift, but the 
statistics are still relatively poor.
The question naturally arises whether or not the space density drop toward
higher $z$ also signifies a change in cardinal properties of the quasar
population at that epoch, given the relatively short time available to
generate massive black holes; at $z \gsim 3$ the Universe was only 
$<$10\% of its present age for a $\Lambda = 0$ Universe
(and $\lsim$15\% of its present age for $\Lambda = 0.7$).

Recent developments allow virial black-hole masses in distant quasars to be 
estimated using scaling relationships calibrated to reverberation 
mapping masses of nearby AGNs and quasars (Wandel, Peterson, \& Malkan 1999;
Kaspi \et 2000; Vestergaard 2002, hereafter Paper~I). 
In this work, these scaling relationships are used to estimate the
black-hole masses in high-$z$ quasars.
The main focus in this work is to compare the typical central masses of 
quasars just below and above the epoch $z \approx 3.5$ at which the 
comoving space density of quasars declines. In addition,   
these masses are discussed in light of our current knowledge of the nature
of their host galaxies.  Representative quasar samples in the redshift range 
$1.5 < z < 6.3$ are studied (\S~\ref{data}), including part of the large, 
high-quality data bases of high-redshift quasars now available.
Several surveys have in the past $\sim$10 years turned up a large 
number of previously unknown $z \approx 4$ quasars (\eg Kennefick 
\et 1995a, 1995b, 1996; Storrie-Lombardi \et 1996; Schneider, Schmidt, 
\& Gunn 1997; see also Constantin \et 2002).
The Sloan Digital Sky Survey (SDSS) has in recent years had particular
success in uncovering an even larger number of quasars at redshifts
greater than 3.5 (Fan \et 1999, 2000, 2001b, 2001c; Anderson \et 2001).
For a large subset of these quasars for which spectra are published,
the black-hole masses were estimated in this work and compared with mass
estimates of lower-redshift quasars, namely the well-selected $z \approx 2$
quasar sample of Vestergaard (2000).  Given current and earlier survey 
capabilities, these samples necessarily consist of luminous quasars.
Since data on nearby luminous quasars and AGNs readily exist in the Bright 
Quasar Survey sample (hereafter BQS; Schmidt \& Green 1983), 
this sample was also included for comparison and reference. 
The lack of objects between $0.5 \leq z \leq 1.5$ does not affect the 
conclusions of this study.

The structure of this paper is as follows.  The approach in estimating \mbh, 
\lbol, and \lol{} is outlined next (\S~\ref{mlmsmts}).  The data are described 
in \S~\ref{data}, presented in detail in \S~\ref{distributions}, and 
discussed in \S~\ref{bhprops}. 
In particular, since the central masses are very large ($\sim 10^9$\Msol{})
for the $z \gsim 4$ quasars, the efficacy of the scaling relations is 
examined in \S~\ref{mreliability}; it is concluded that the scaling 
relations are fully applicable to the high-$z$ quasars with an uncertainty 
of a factor $\lsim$4.
Section~\ref{qhosts} discusses the fact that although the masses of these
high-$z$ active black holes are rather large, the AGN host galaxies at 
$z \gsim 4$ appear to be rather young, and even small, young galaxies
at $z \approx 3$ are capable of hosting massive black holes.        
This leads to the conclusion that super-massive black holes were in place
long before the bulk of the stellar mass in the quasar host galaxies.
A summary and conclusions are provided in \S~\ref{conclusion}.

A cosmology with $H_0$ = 75 ${\rm km~ s^{-1} Mpc^{-1}}$, q$_0$ = 0.5, and 
$\Lambda$ = 0 is used throughout, although the main conclusions of this study 
do not depend on the specific choice of cosmology. 
However, it is worth mentioning that adoption of a cosmology of $H_0$ = 70 
${\rm km~ s^{-1} Mpc^{-1}}$, $\Omega_M$ = 0.3, $\Omega_{\Lambda}$ = 0.7 will 
increase the mass estimates at the highest redshifts by no more than a factor 
2.4 and by a decreasing factor at lower redshifts.
%

\vspace{0.7cm}
\section{M$_{\uppercase{\rm BH}}$, L$_{\rm bol}$, and L$_{\rm bol}$/L$_{\rm {\uppercase{E}}dd}$ Measurements \label{mlmsmts}}

The samples of quasars studied here cover the redshift
ranges $z \leq 0.5$ and $1.5 \lsim z \lsim 6.3$. Therefore, the observed optical 
spectra cover restframe wavelengths including \hb{} for the BQS and the region 
around \civ{} for the higher redshift quasars.

The virial masses, $M \propto v^2 R$, are estimated from single-epoch spectral
measurements using the relationships presented in Paper~I (eqns.\ 1, 2, A5, and 7), 
since they are relevant for spectroscopic data covering the wavelength regions 
around the \hb{} and \civ{} lines.  Here $v$ is the Doppler width of an emission 
line and $R$ is the radius of the line-emitting region.
Note that also the high-ionization line-emitting gas, including \civ{}, 
yields virial masses consistent with those derived from \hb{} in
all the objects for which this can be tested (Peterson \& Wandel 2000; 
Onken \& Peterson 2002). Hence, the \civ{} line width is valid 
as a measure of $v$ (see discussion in \S~\ref{mreliability}).

The important scaling relationship for the mass equations listed below is 
the radius $-$ luminosity relation, that allows the size of the broad line 
region (BLR) to be estimated from the optical continuum luminosity of the 
source.  The empirical scaling law, applicable to Balmer lines and based on 
reverberation-mapped nearby AGNs and quasars, is (Kaspi \et 2000): 
\begin{equation}
R_{\rm BLR} = (32.9\pm 5.5)~\left[ \frac{ \lambda L_{\lambda}
\rm (5100\AA) }{ 10^{44}~\rm ~ergs\,s^{-1}} \right]^{0.7\,\pm\,0.1}~~{\rm
lt-days.}
\label{blrsize.eq}
\end{equation}
The mass equation based on optical data (hereafter, the `optical relationship') is 
\vskip 0.2cm \noindent
\begin{eqnarray}
\lefteqn{\log \,M_{\rm BH}/\Msol = \mbox{~~~} 6.7 ~~+ }\\
\nonumber \\
    & & \log \,\left[ \left({\rm \frac{FWHM(H\beta)}{1000~km~s^{-1}}} \right)^2 ~
    \left( \frac{\lambda L_{\lambda} \rm (5100\AA)}{10^{44} \rm 
     ~ergs~s^{-1}}\right)^{0.7} \right] \nonumber 
\label{mrl.eq}
\end{eqnarray}
\vskip 0.2cm \noindent
and the mass equation relevant for UV data (the `UV relationship') is

\begin{eqnarray}
\lefteqn{\log \,M_{\rm BH}/\Msol =  \mbox{~~~} 6.2 ~~+ } \\ 
\nonumber \\
   & &  \log \,\left[ \left({\rm \frac{FWHM(C\,IV)}{1000~km~s^{-1}}} \right)^2 ~
   \left( \frac{\lambda L_{\lambda} \rm (1350\AA)}{10^{44} \rm ~ergs~s^{-
1}}\right)^{0.7} \right], 
\label{logmuv.eq}
\end{eqnarray}
\vskip 0.3cm \noindent
where the sample standard deviation of the weighted mean is 0.45 and shows the 
intrinsic scatter in the sample (Paper~I).
This is confirmed by the black-hole masses of reverberation mapped AGNs 
which exhibit an intrinsic scatter 
of about 0.3 $-$ 0.5\,dex (Peterson 2002) around the $M_{\rm BH} - \sigma$ 
relationship, established by local quiescent galaxies (\eg Ferrarese \& Merritt 
2000; Gebhardt \et 2000a; see also Gebhardt \et 2000b and Ferrarese \et 2001 
for comparisons of some AGN masses with the $M_{\rm BH} - \sigma$ relation).
Both the optical and the UV relationships were calibrated (Paper~I) to virial mass 
estimates of low-redshift active galaxies determined from reverberation-mapping 
analysis.

The velocity dispersion, $v$, of the broad line gas is estimated using the broad 
emission line width as characterized by FWHM. While equation~(\ref{blrsize.eq})
is valid for the Balmer lines only, a similar scaling law, $R_{\rm BLR} \propto L^{0.7}$,
is assumed for the UV relationship; the constant of proportionality is determined
by the calibration and included in equation~(\ref{logmuv.eq}).
Note that when applying these scaling relationships to higher-$z$ quasars, a
large extrapolation in luminosity is not performed, which might otherwise
raise concern. The efficacy of the mass scaling relationships is 
discussed in \S~\ref{mreliability}.

The bolometric luminosities are estimated based
on existing average spectral energy distributions (SEDs).
SEDs are only available for a small group of distant quasars. Kuhn \et (2001) 
found no significant differences between the restframe optical and UV   
energy distributions for 15 $z > 3$ quasars and low-$z$ quasar SEDs. 
It is fair to assume the Kuhn \et sample is representative, also of
quasars at higher redshifts yet, given the similarity of quasar spectra 
at all redshifts (see \S~\ref{bhprops}). Specifically, this means that 
the empirical average SED of low-$z$ active galaxies and quasars (Elvis \et 
1994) is also representative of the average SED of more distant quasars. 
Here, the average radio-quiet quasar (RQQ) SED of Elvis \et (1994) is updated 
to reflect the more representative optical-to-X-ray slope of 1.43 determined 
by Elvis, Risaliti, \& Zamorani (2002) using new and improved X-ray data.
This updated RQQ SED was used to estimate bolometric correction factors 
applicable also to radio-loud quasars (RLQs), because (1) high-$z$ RLQs 
appear to constitute a similar low fraction ($\sim$10\%) of all quasars 
as seen for $z \lsim$ 2 quasars (Stern \et 2000), and (2) the average 
RLQ SED has not yet been updated (M.\ Elvis, 2002, private communication).
The bolometric correction factors to the monochromatic continuum luminosities, 
$\lambda L_{\lambda}$, at rest wavelengths of 1350\AA, 1450\AA, and 4400\AA{} 
were computed\footnote{ 
The bolometric correction to $\lambda L_{\lambda}$(4400\AA) of 9.74 is slightly 
lower than the average correction of 11.8 established by Elvis \et (1994). 
The reason is in part the weaker contribution from the X-rays, and in part 
due to the use of the average SED in this work compared to the average of the 
individual bolometric correction factors quoted by Elvis \et (1994). The 
scaling factor of 9.74 is preferred for $\lambda L_{\lambda}$(4400\AA) for 
consistency with the 1350\AA{} and 1450\AA{} bolometric correction factors 
which were not quoted by Elvis et al.} 
to be 4.62, 4.65, and 9.74, respectively.  
Applying a bolometric correction to the UV continuum luminosities for the
high-$z$ quasars reduces the uncertainty in the $L_{\rm bol}$ estimate 
compared to using a bolometric correction to luminosities at other wavelengths. 
The reasons are that one eliminates the uncertainty associated with the 
measured UV continuum slope and extrapolation to other wavelengths, and
since $\lambda$1350 and $\lambda$1450 are close to the peak of the big blue 
bump, the associated bolometric correction factors are the smallest relative
to other wavelengths. 

The conservative measurement uncertainties estimated from the line widths,
continuum fluxes, and continuum slopes were propagated to estimate the 
measurement errors on \mbh, $L_{\rm bol}$, and the Eddington ratios 
$L_{\rm bol}/L_{\rm Edd} \propto L_{\rm bol}$/\mbh{}. These
uncertainties do not include the uncertainty associated with the use of
bolometric corrections or the use of the simple scaling relationships in 
eqn.~(\ref{mrl.eq}),~(\ref{blrsize.eq}), and~(\ref{logmuv.eq}) to 
estimate the \mbh. A statistical 1$\sigma$ uncertainty of 
about 0.4\,dex and 0.5\,dex (\ie a factor 2.5 and 3 in the mass) 
for the optical and UV relationships, 
respectively, was found for the sample studied in Paper~I. 
This additional uncertainty must be kept in mind, especially when considering
the masses in absolute terms. However, since \mbh{} can likewise be
reasonably estimated for high redshift quasars (\S~\ref{mreliability}), 
statistical analysis and 
comparison of the relative masses of high-redshift quasars should be 
reasonably sound.

\section{Data \label{data}}

Several samples of quasars at low, intermediate, and high redshift
are analyzed here as described in detail below.

\subsection{Low Redshift Quasars 
\label{bqsdata}}

This sample consists of the 87 BQS quasars at $z \leq 0.5$.
For the objects with available reverberation-mapping masses, these
masses are used, as computed in Paper~I. For the remaining sources, 
the masses are estimated from equation~(\ref{mrl.eq}).
The measurements of line widths and continuum luminosities used are 
explained in detail in Paper~I.  
Briefly, the FWHM(\hb) measurements were primarily adopted from Boroson
\& Green (1992).
The monochromatic luminosities, $L_{\lambda}$\,(5100\AA), were determined for
most of the objects from the spectrophotometry by Neugebauer \et (1987),
but for a few sources $L_{\lambda}$(5100\AA) was computed from 4400\,\AA{} flux 
densities (Kellerman \et 1989) by adopting a power-law slope, 
$\alpha ~=~-$0.5 (F$_{\nu} \propto \nu^{\alpha}$).
Bolometric luminosities for different subsets of the BQS quasars have
been measured by a few authors (Sanders \et 1989; Elvis \et 1994; 
Lonsdale, Smith, \& Lonsdale 1995; Wilkes \et 1999a). Comparison of the
measurements for quasars common between two or more of these studies shows
that the \lbol{} values are generally relatively similar\footnote{While one
can argue that one of the studies has better X-ray data, then another has
better IR data, and for individual objects, the SED of one study may be
preferred over that of another. However, it is beyond the scope of this 
work to compile updated SEDs based on the best data now available.} 
within a common scatter of 0.2 $-$ 0.3\,dex, consistent with the estimated 
measurement uncertainties quoted by Elvis \et (1994). In the interest of
simplicity and consistency with the data used for the quasar samples at
higher-redshifts, the \lbol{} values were here estimated as $(9.74\pm 4.3) 
\times \lambda L_{\lambda}$(4400\AA) (\S~\ref{mlmsmts}). Adopting this scaling
introduces an uncertainty similar to the measurement uncertainties quoted
above and is thus considered acceptable.
Errors on the parameters are propagated using the measurement uncertainties
on the primary measurements as described in Paper~I.

\subsection{Intermediate Redshift Quasars \label{thesisdata}}

The sample of 114 quasars at intermediate redshifts ($1.5 \lsim z \lsim 3.5$)
studied here is that analyzed by Vestergaard (2000).
The sample consists of 68 RLQs and 48 RQQs.  
The RLQs were selected from Barthel \et (1988), Barthel, 
Tytler, \& Thomson (1990), Murphy, Browne, \& Perley (1993), and the 
3C and 4C radio catalogs to span the range of observed quasar radio 
spectral indices; this allows a study of how the spectral features may change 
with source inclination with respect to our line of sight (\eg Vestergaard,
Wilkes, \& Barthel 2000).  Most of the RLQ data are from Barthel \et (1990).
The individual RQQs were selected from Hewitt \& Burbidge (1993)
to be well-matched and paired in redshift and luminosity to individual quasars
in the radio-loud subset. The purpose was to study the rest 
UV spectral characteristics with radio properties (Vestergaard 2000).
Given the careful sample selection, this sample of 116 quasars is here 
considered reasonably representative of the $z \approx 2$ quasar population 
to be compared with the higher redshift quasars. The full sample, as opposed 
to the pair-matched subset, is studied here to improve the statistical
significance of the comparison.

The \civ{} line width, FWHM(\civ), was measured reliably for this sample
on a smooth representation of the emission profile, obtained by modeling
with 2 to 3 Gaussian functions (following Laor \et 1994), above a global
power-law continuum fit to line-free windows (\eg Francis \et 1991; 
Vestergaard \& Wilkes 2001). The FWHM uncertainties were estimated by
varying the continuum fit within the continuum flux $rms$ and remeasuring
the FWHM. 
The continuum fluxes were measured from the best fit power-law and the flux 
uncertainties were determined as the $rms$ scatter around this fit in the 
continuum fitting windows.  
The bolometric luminosity is estimated from $L_{\lambda}$(1350\AA), \ie 
\lbol = 4.62 $\times \lambda L_{\lambda}$(1350\AA) [\S~\ref{mlmsmts}].

\subsection{High Redshift Quasars
\label{hizdata}}

The high-$z$ quasar sample consists of $\sim$150 $z > 3.5$ quasars, 
mostly from the SDSS,
selected from recently published work, as detailed below. 
For consistency, the spectra of these quasars were measured in a similar 
manner, as follows: 
FWHM(\civ) was measured relative to a reasonable, best-fit continuum 
determined using virtually line-free continuum windows. 
The continuum flux at rest frame wavelengths 1350\AA{} and 1450\AA{} 
were measured from this best-fit power-law continuum.
Conservative measurement uncertainties on line widths, continuum fluxes, 
and spectral slopes were estimated based on the continuum {\it rms} noise, 
as described above for the intermediate-$z$ quasars.
The spectral slope measured here is only used when more accurate,
published values are unavailable.  \lbol{} was estimated as 
4.65 $\times \lambda L_{\lambda}$(1450\AA) (see \S~\ref{mlmsmts}).

The source studies from which the individual quasars are drawn are
listed below. In each case, the details of the analysis specific to the 
individual data set are outlined. 

\paragraph{Constantin \et (2002):}
High-quality spectra of these $z \approx 4$ quasars were kindly 
provided by the authors.
Constantin et al. corrected their data for Galactic reddening using 
Schlegel, Finkbeiner, \& Davis (1998) $A_V$ values and the Cardelli, 
Clayton, \& Mathis (1989) extinction curve.  Constantin et al.\ 
quote an uncertainty of 0.2 dex in their AB(1450\AA)-based, 
monochromatic luminosities.  The $L_{\lambda}$(1450\AA) values measured 
here were recalibrated to match the AB-based $L_{\lambda}$(1450\AA) 
for the few objects showing a difference larger than 0.2\,dex between 
the two sets of luminosity measurements.
The spectra of 40 of the 44 quasars could be measured. 
A few quasars are too strongly absorbed in the \civ{} emission profile 
for a reasonable measurement of the line width. One quasar exhibited
no conspicuous emission lines other than \lya{} and \nv.

\paragraph{Anderson \et (2001):}
The digital SDSS spectra of these $\sim$125 quasars are publicly available.
The best quality spectra were selected for measurement.  A few other poorer 
quality spectra were also measured. The spectra were first corrected
for Galactic reddening using the Schlegel \et (1998) $E(B-V)$ values obtained 
from NED and the Cardelli \et (1989) reddening curve. 
For a subset ($\sim$60) of the quasars Fan \et (1999, 2000, 2001b) present 
AB(1450\AA) magnitudes to which the spectral measurements were recalibrated.

\paragraph{Fan \et (1999, 2000, 2001b):} The best-quality spectra (kindly 
provided by X.\ Fan) of these SDSS quasars that include a 
full, measurable \civ{} line profile were selected for measurement. 
The continuum luminosities measured here at 1350\AA{} and 1450\AA{} 
were recalibrated to the (dereddened) AB(1450\AA) magnitudes.  
Fan \et define 3 quasar subsamples. 
The set of 59 quasars analyzed here (hereafter the ``Fan \et sample'')
is drawn from these subsamples, as follows: 
(a) 34 of the 39 quasars that comprise the well-defined ``color-selected 
sample'' (3.6 $\lsim z \lsim$ 5; $i^{\ast} \leq  20$ mag), complete within 
the survey area (see Fan \et 2001b for details),
(b) 12 quasars from the ``bright subsample'' ($i^{\ast} \lsim 20.2$ mag)
that are not included in (a), 
and (c) 13 of the 18 ``faint subsample'' quasars ($i^{\ast} < 21$ mag). 
As mentioned, the selection is mostly based on data quality (but see below). 
Ten quasars were common between the originally selected ``Anderson \et'' and the 
``Fan \et'' samples.  Since neither of these samples are complete, each of 
those quasars only occur once in the ``high-$z$ sample''; each object was 
included in the sample which has the better spectrum.
The data of the ``color-selected quasars'' are presented by
Fan \et (1999, 2001b) and Schneider \et (2001).
The photometry of Fan \et (2001b) and line widths measured in the high 
$S/N$ spectra of Constantin et al. were used for two objects common 
to these studies. Five quasars were excluded from the ``color-selected
sample'': the spectra of three of these quasars did not include the 
\civ{} emission line.  For another object (J020731.68$+$010348.9) the 
poor spectrum quality precludes a FWHM(\civ) measurement.  Finally, the 
spectrum of a fifth object, J225529.09$-$003433.4, could not be located 
in any of the references listed or in the SDSS archive. 
It will be shown in the next section that the ``color-selected sample''
is statistically similar in the distributions of \lbol, \mbh, and \lol{} 
to those of the other high-$z$ quasar samples ($3.5 < z \lsim 5.5$), as 
expected \mbox{if this is a representative subset of $z \approx 4$ quasars.}


\begin{figure*}[th]
\vbox{
\hbox{
\epsfxsize=6.0cm
\epsfbox{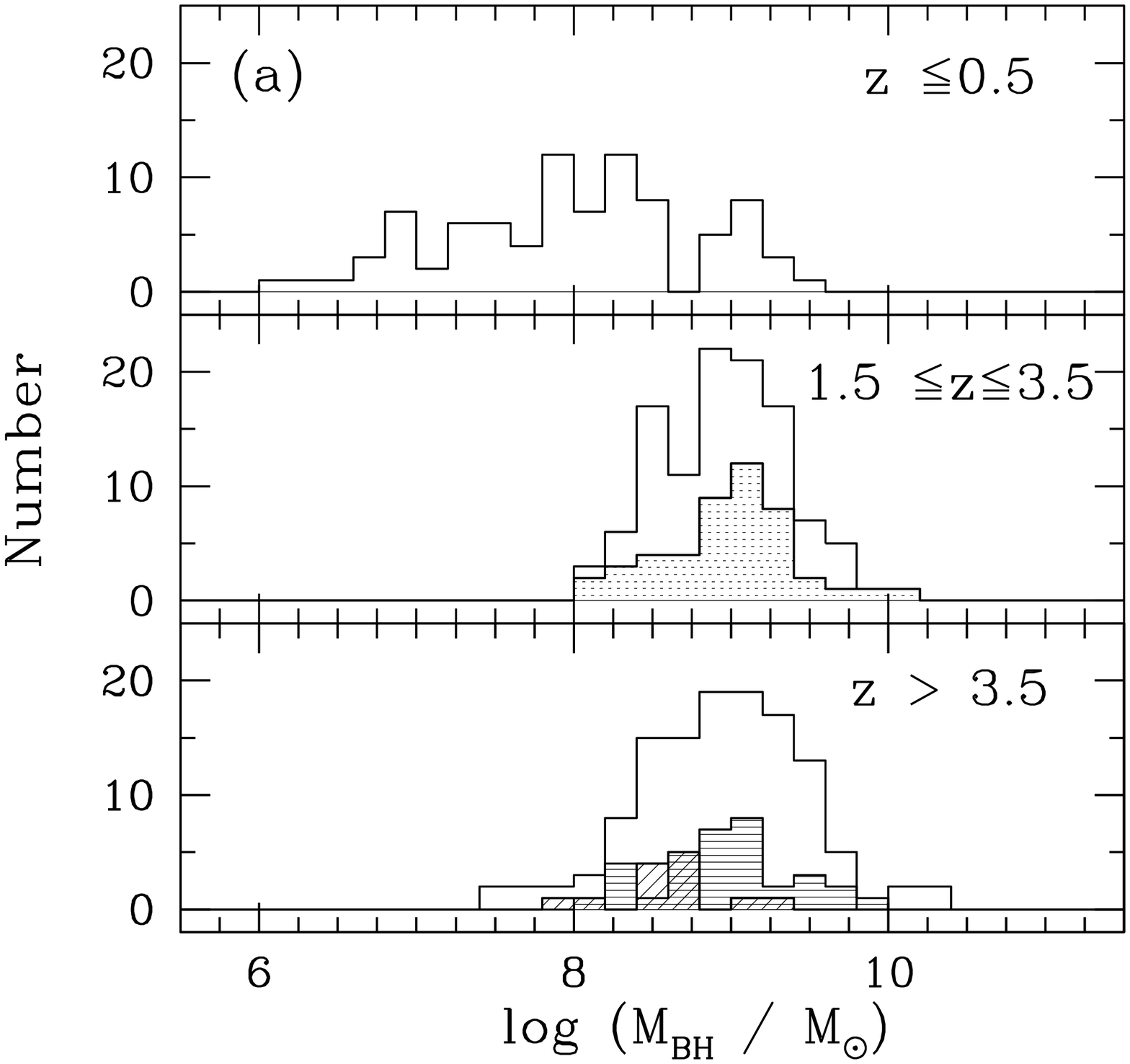}
}\vspace{-6.0cm}\hspace{6.25cm}\hbox{
\epsfxsize=6.0cm
\epsfbox{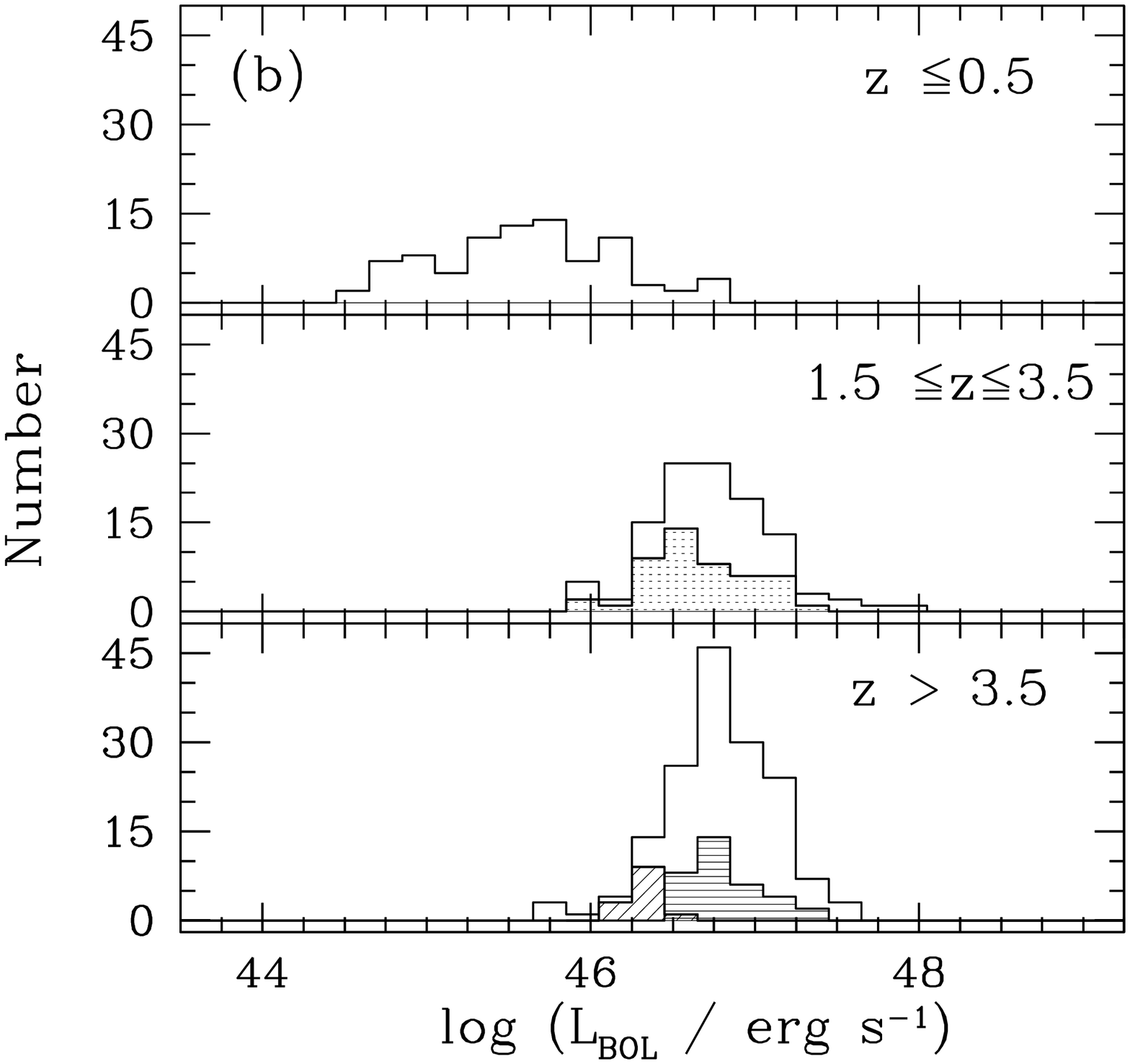}
}\hbox{
\epsfxsize=6.0cm
\epsfbox{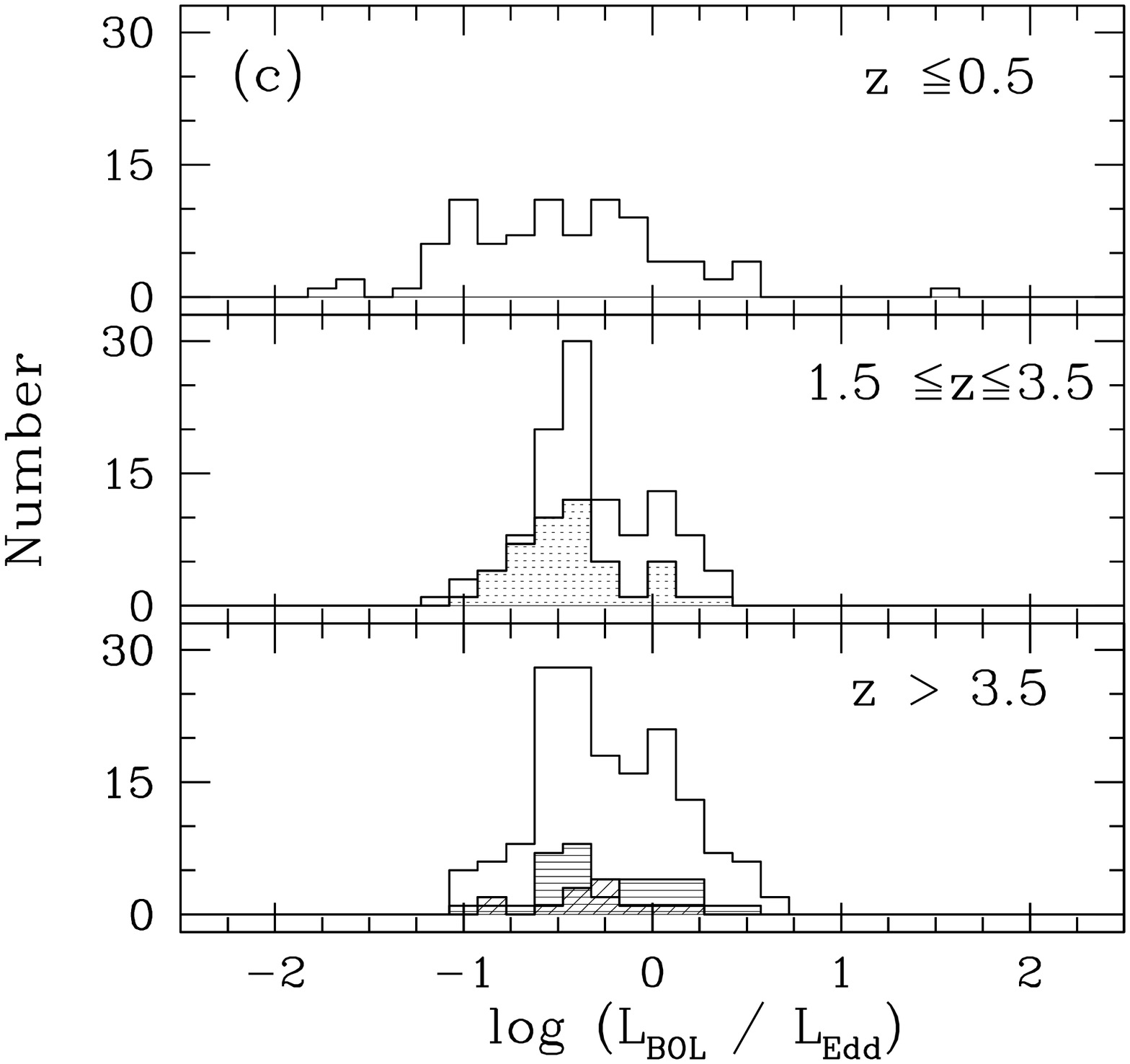}
}
}
\begin{center}
\caption[]{ 
Distributions of estimated ($a$) black-hole mass, \mbh, ($b$) bolometric luminosity, \lbol, 
and ($c$) Eddington luminosity ratio, \lol{} for different redshift bins.
The top panel consists of Bright Quasar Survey objects only.
The middle panel contains the intermediate-$z$ sample; the radio-quiet quasars
(RQQs) are shown shaded. The bottom panel unshaded histogram contains all the
($z > 3.5$) quasars in the samples described in \S~\ref{hizdata}. 
The distribution of the faint subsample of Fan \et (2001b) is high-lighted with 
diagonal shading. The histogram for the ``color-selected sample'' of Fan \et
(2001b) is shown with a light horizontal shading for comparison.
\label{mlhistzbins.fig}}
\end{center}
\end{figure*}


\begin{figure*}[]
\begin{center}
\vbox{
\hbox{
\hspace{1.3cm}
\epsfxsize=7.0cm
\epsfbox{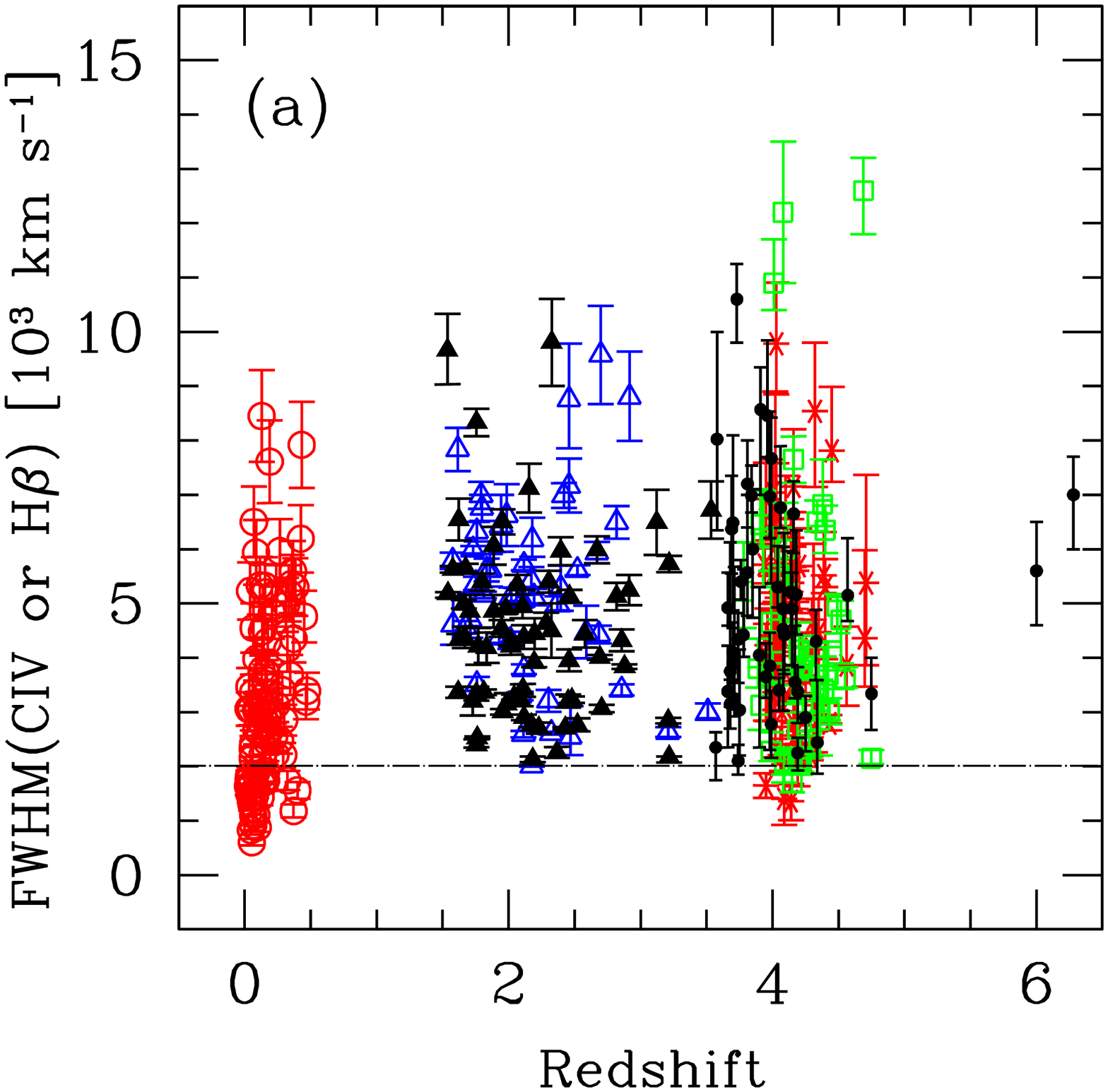} }
\vspace{-7.0cm}\hspace{7.0cm}
\hbox{
\epsfxsize=7.0cm
\epsfbox{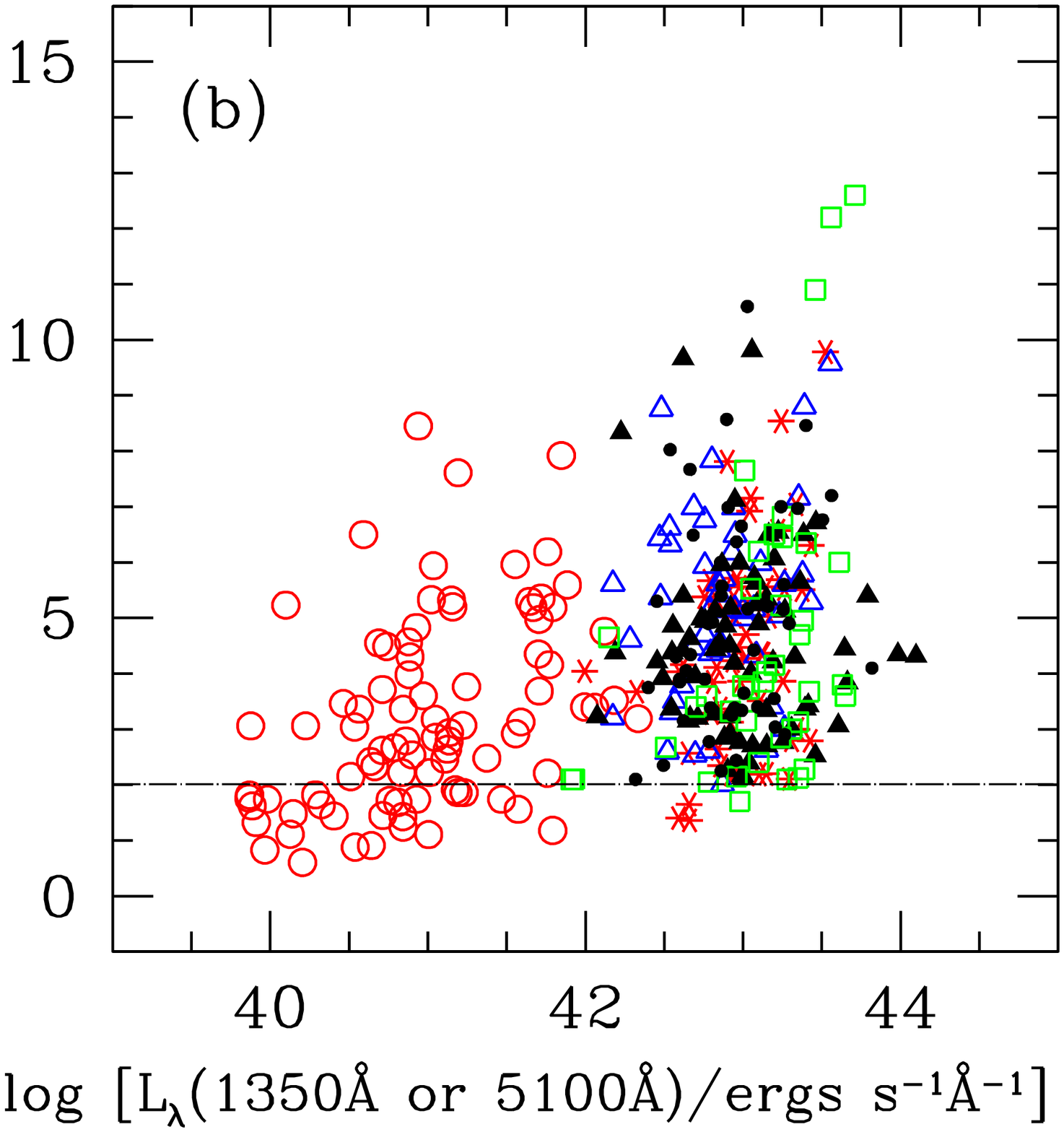}
}}
\caption[]{Distributions of FWHM(\hb) (for BQS; open circles) and FWHM(\civ)
(remaining quasars) with ({\it a}) redshift and ({\it b}) continuum luminosity,
$L_{\lambda}$(5100\AA) (for BQS) and $L_{\lambda}$(1350\AA) (for the $z > 1$
quasars). The horizontal, dashed line marks FWHM = 2000 \kms{}, the commonly
adopted cutoff for selecting quasars.
Samples shown are: BQS (open circles); intermediate redshift radio-quiet (open
triangles) and radio-loud (solid triangles) quasars; Constantin \et sample
(open squares); Anderson \et sample (asterisks); Fan \et sample and
$z \approx 6$ quasars (solid points).
\label{fw_lum.fig}}
\end{center}
\vspace{-0.50cm}
\end{figure*}

\paragraph{Pentericci \et (2002):} 
The authors present near-IR spectra of two of the $z \approx 6$ 
quasars discovered in SDSS (Fan \et 2001c), namely SDSSp J103027.10$+$052455.0 
($z$ = 6.28) and SDSSp J130608.26$+$035626.3 ($z$ = 6.00). The \civ{}
line widths are estimated from enlarged versions of the published spectral
plots, while the luminosities are adopted from Fan \et (2001c). 
The width of \civ{} measured here for SDSSp J103027.10$+$052455.0 is 
consistent with that quoted by Pentericci \et within the errors.
The \civ{} profile is truncated in the blue-most part of the wing.
The FWHM estimate (=7000$^{+700}_{-1000}$ \kms) used here for \mbh{} was 
obtained by measuring a reasonable extension of the observed blue profile 
to the continuum level indicated. The (large) FWHM measurement uncertainty 
adopted is based on a conservative estimate of the continuum uncertainty.

\section{Distributions of Mass, Luminosity, and Eddington Ratio \label{distributions}}


\begin{figure*}[t]
\begin{center}
\vbox{
\hbox{
\hspace{1.3cm}
\epsfxsize=7.0cm
\epsfbox{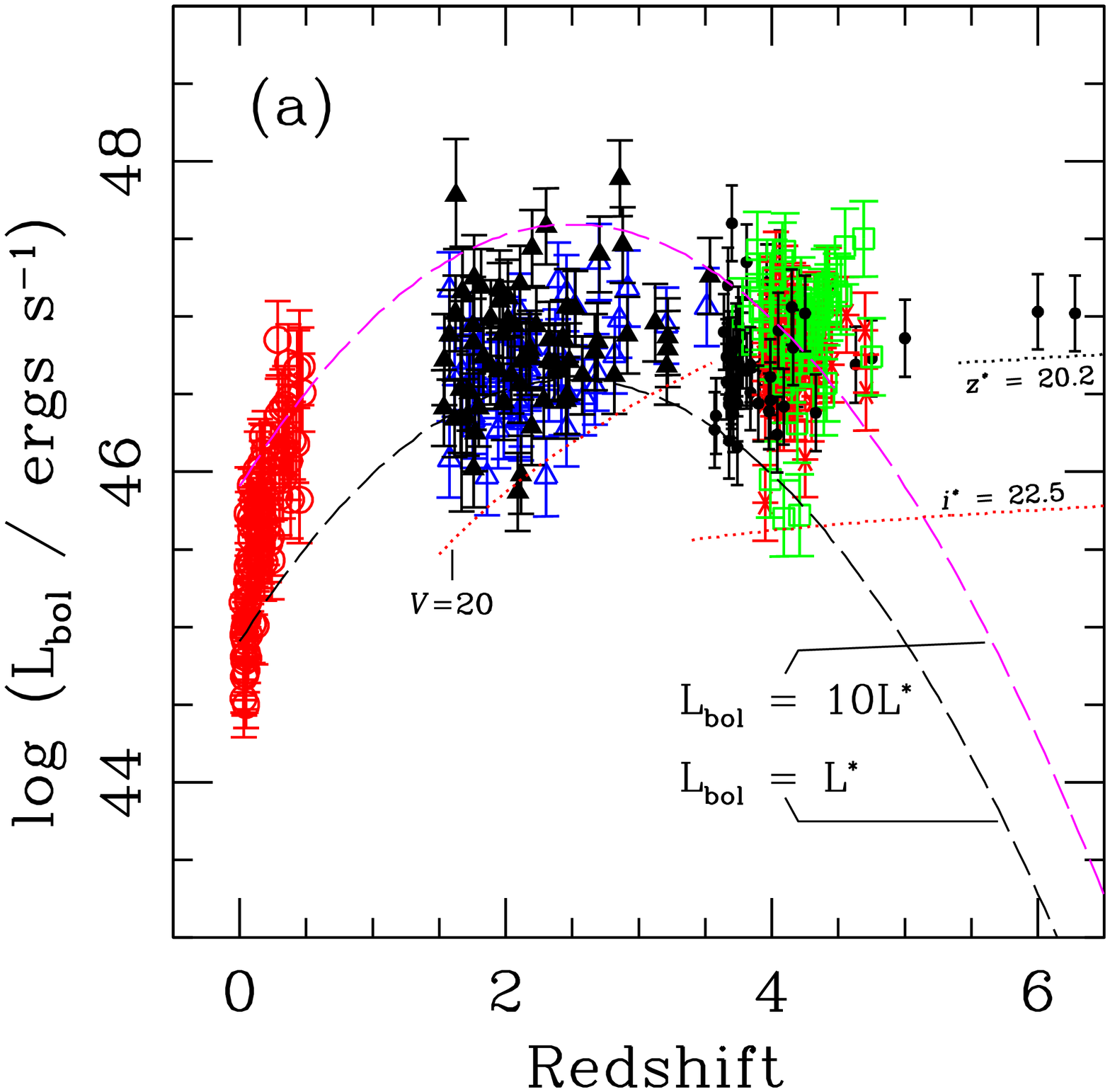} }
\vspace{-7.0cm}\hspace{7.0cm}
\hbox{
\epsfxsize=7.0cm
\epsfbox{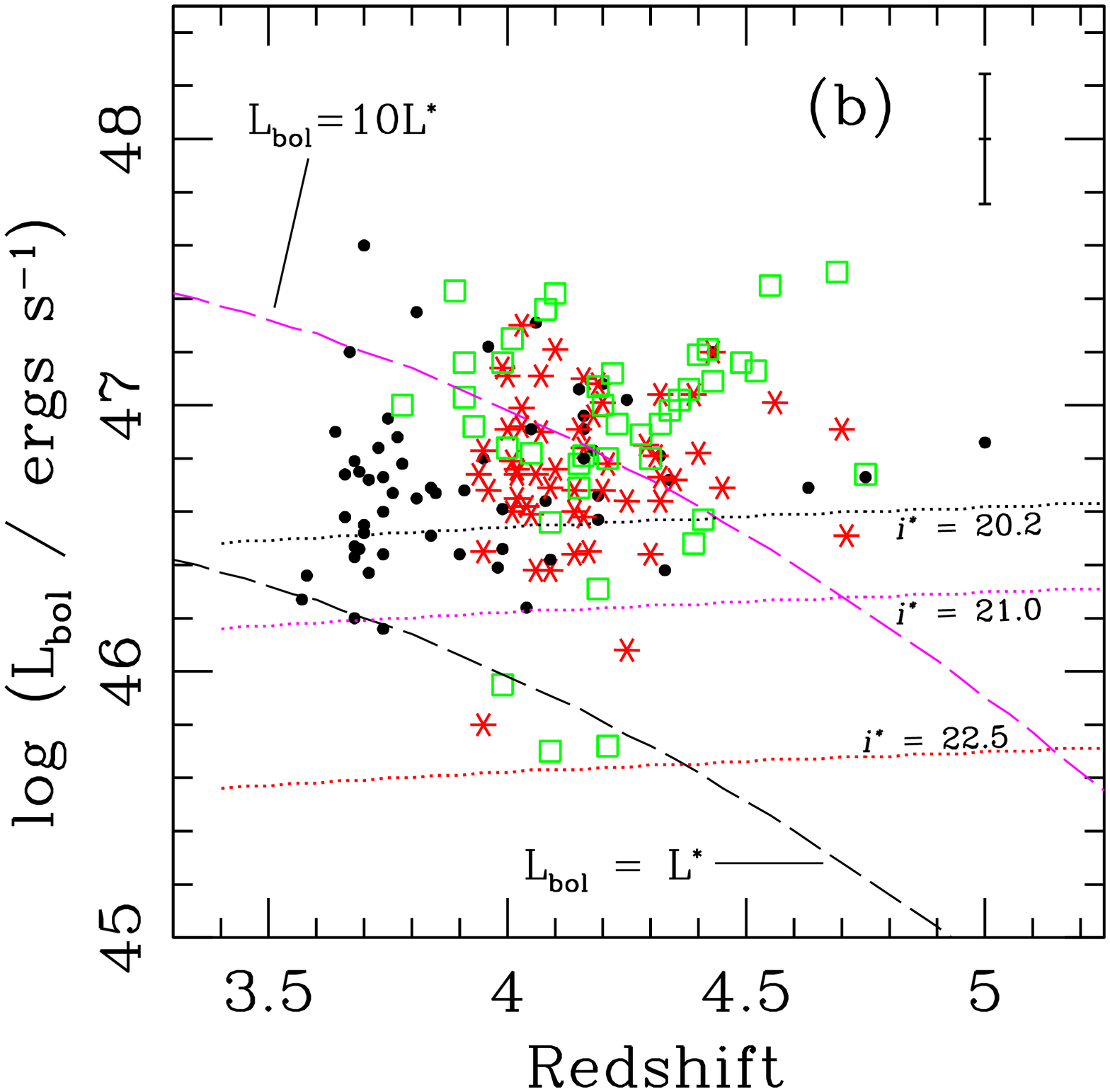}
}}
%
\caption[]{Distribution of bolometric luminosities, \lbol, as a function
of redshift for ({\it a}) all the quasar samples considered here and
({\it b}) the $z > 3.6$ quasar samples only for clarity; error bars are
omitted but the average error is shown in upper right corner.
See Fig.~\ref{fw_lum.fig} for symbols.
The dashed curves represents the evolution of \lbolstar{}(\lstar) (lower)
and \lbolstar{}(10\,\lstar) (upper) with redshift. The
evolution of \lbolstar{} at $z \gsim 2.5$ is an extrapolation of the observed
evolution at lower redshift (see text).
The dotted curve segments denote limits imposed by sample selection and
survey limits. The left panel shows limits for the $z \approx 2$
quasars ($V = 20$ mag), the $z \approx 4$ quasars ($i^{\ast} = 22.5$ mag), and
the $z \approx 6$ quasars ($z^{\ast} = 20.2$ mag).
The right panel shows the limit of the SDSS in general ($i^{\ast} = 22.5$ mag)
and, for reference, the limits of Fan \et's ``bright'' ($i^{\ast} =20.2$ mag)
and ``faint'' ($i^{\ast} =21$ mag) samples.
\label{l_z.fig}}
\end{center}
\end{figure*}

\subsection{Number Distributions}

The histograms of masses, bolometric luminosities, and Eddington ratios are shown 
in Figure~\ref{mlhistzbins.fig} for different redshift bins.
Figure~\ref{fw_lum.fig} shows the distributions of emission-line widths
with redshift and continuum luminosities.

It is evident from Figure~\ref{mlhistzbins.fig} that the intermediate 
and high-redshift ($z > 1.5$) quasars are all massive ($\gsim 10^8$\Msol) 
and luminous ($\gsim 10^{46}$\ergs{}) and have somewhat similar \mbh{} 
and \lbol{} distributions.
As discussed in \S~\ref{zdistr} the lower boundaries are due to selection effects.
The two radio types of the intermediate-redshift quasars, which are matched in 
luminosity and redshift, display no significant \mbh{} difference, in contrast 
to some studies claiming black-holes of RLQs are more massive (\eg Laor 2000;
McLure \& Dunlop 2001). The radio-loud $z \approx 2$ sample does have
a larger number of luminous quasars than the radio-quiet subset (see
\S~\ref{thesisdata}).  Also, even the quasars in the high-redshift 
($z \approx 4$) faint sub-sample of Fan \et (2001b) are quite 
massive ($\sim10^9$\Msol).

Kolmogorov-Smirnov tests confirm the apparent similarity
seen in Figure~\ref{mlhistzbins.fig} of the distributions in mass, 
luminosity, and Eddington ratio for all high-redshift samples (\S~\ref{hizdata}),
including the ``color-selected sample'' of Fan \et (2001b).
The low-redshift quasars of the BQS
span a larger range in each of these parameters
simply because less-luminous quasars are still bright and
easily detectable at low redshift.

Figure~\ref{mlhistzbins.fig}c suggests that in every redshift range 
there is a significant number of quasars with \lol{} greater than unity. 
However, given the uncertainties in both \mbh{} and bolometric corrections,
the reality of
super-Eddington luminosities is unclear. Nevertheless,
barring some large systematic error, all the values of
\lol{} are high, almost always greater than 0.1.
This is not surprising given that we are sampling the
brightest quasars in every redshift bin. Indeed,
the lower end of the distribution in \lol{} is
due to survey flux limits and the apparent upper envelope to the \civ{} line
width for a given luminosity (Fig 2b).
The upper end of the distribution is truncated by the lack of objects with
very high luminosity and/or emission lines narrower than about 2000\,\kms{}.
Since active nuclei are commonly identified by their broad lines, it
is not clear whether the discovery surveys with their selection criteria
(and crude spectral resolution in early surveys) would be able to identify
quasars with such narrow lines; even the SDSS quasar identification is not
well-defined at present (Richards \et 2001).
The upper cutoff in \lbol{} is however real (see e.g., \S~\ref{zdistr}).

\vskip 0.6cm \noindent
\subsection{Redshift Distributions \label{zdistr}}


\begin{figure*}[t]
\begin{center}
\vbox{
\hbox{
\hspace{1.3cm}
\epsfxsize=7.0cm
\epsfbox{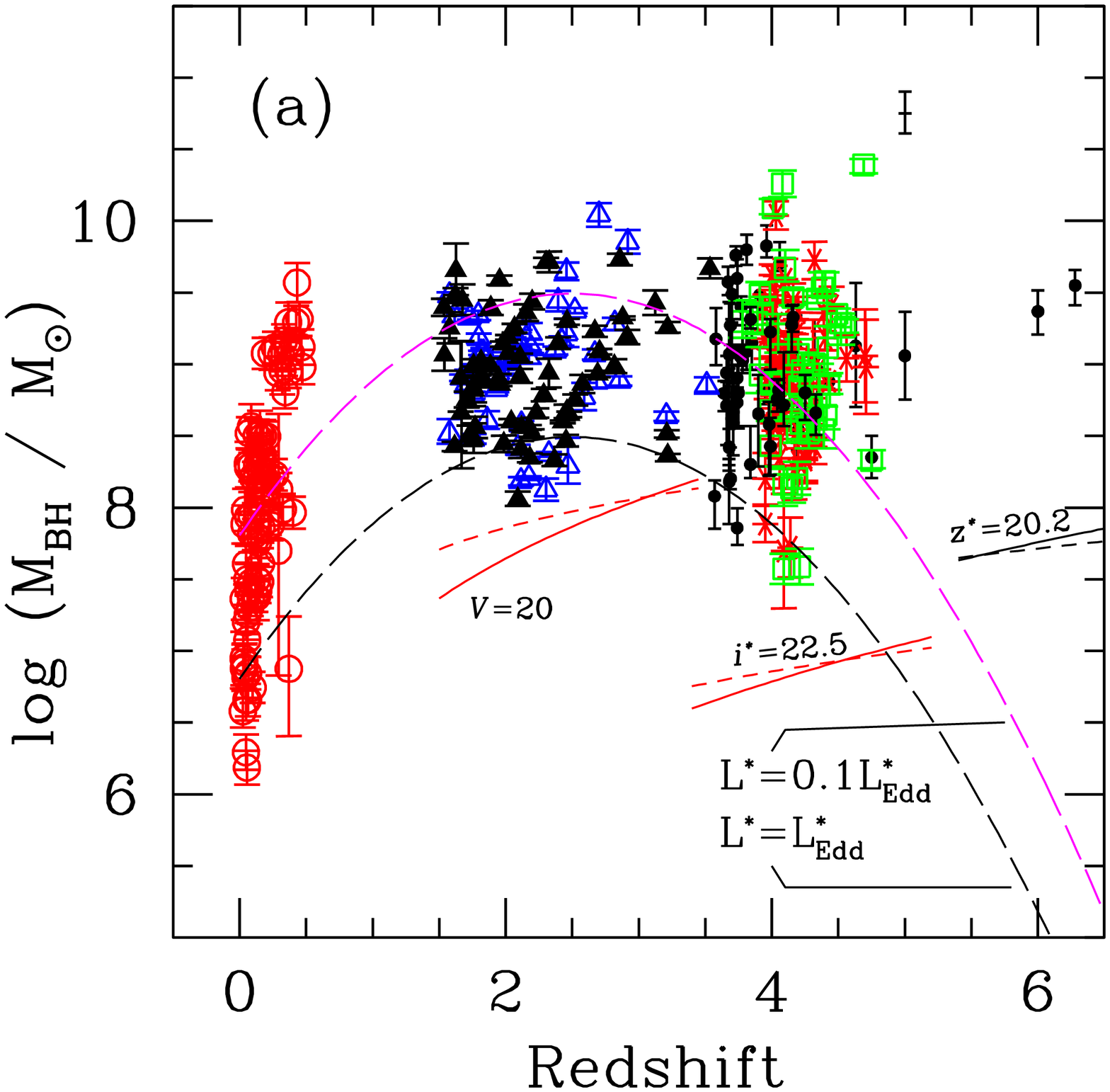} }
\vspace{-7.0cm}\hspace{7.0cm}
\hbox{
\epsfxsize=7.0cm
\epsfbox{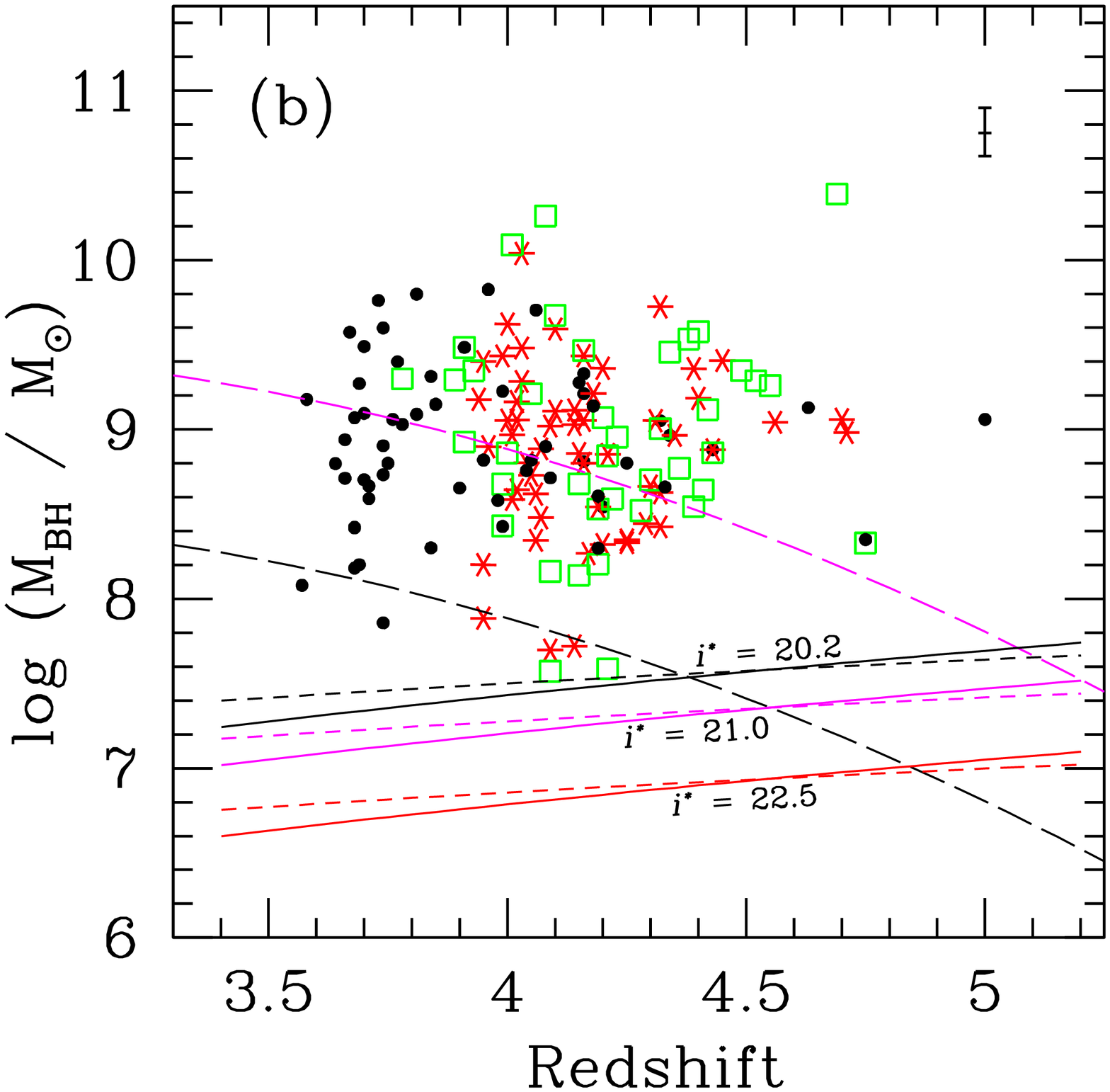}
}}
\caption[]{Distributions of \mbh{} with redshift for ({\it a}) all samples
considered here, and ({\it b}) the $3.5 \leq z \leq 5.0$ samples only for
clarity; the average error is shown in upper right corner.
See Fig.~\ref{l_z.fig} for symbols.
The dashed curves show the observed ($z \lsim 2.5$) and predicted ($z > 2.5$)
evolution of \mstar{} determined by assuming \lbolstar = \leddstar{} (lower)
and \lbolstar = 0.1 \leddstar{} (upper), respectively. The line segments denote
limits in \mbh{} imposed by the survey flux limits (see Fig.~\ref{l_z.fig})
assuming FWHM(\civ) = 2000\,\kms{} (1000\,\kms) for $z \approx$ 2 ($z \gsim$4)
quasars, and adopting the average (dashed segment) and minimum (solid segment)
spectral slopes for the quasars at each redshift interval.
\label{m_z.fig}}
\end{center}
\end{figure*}

\begin{figure*}[t]
\begin{center}
\vbox{
\hbox{
\hspace{1.3cm}
\epsfxsize=7.0cm
\epsfbox{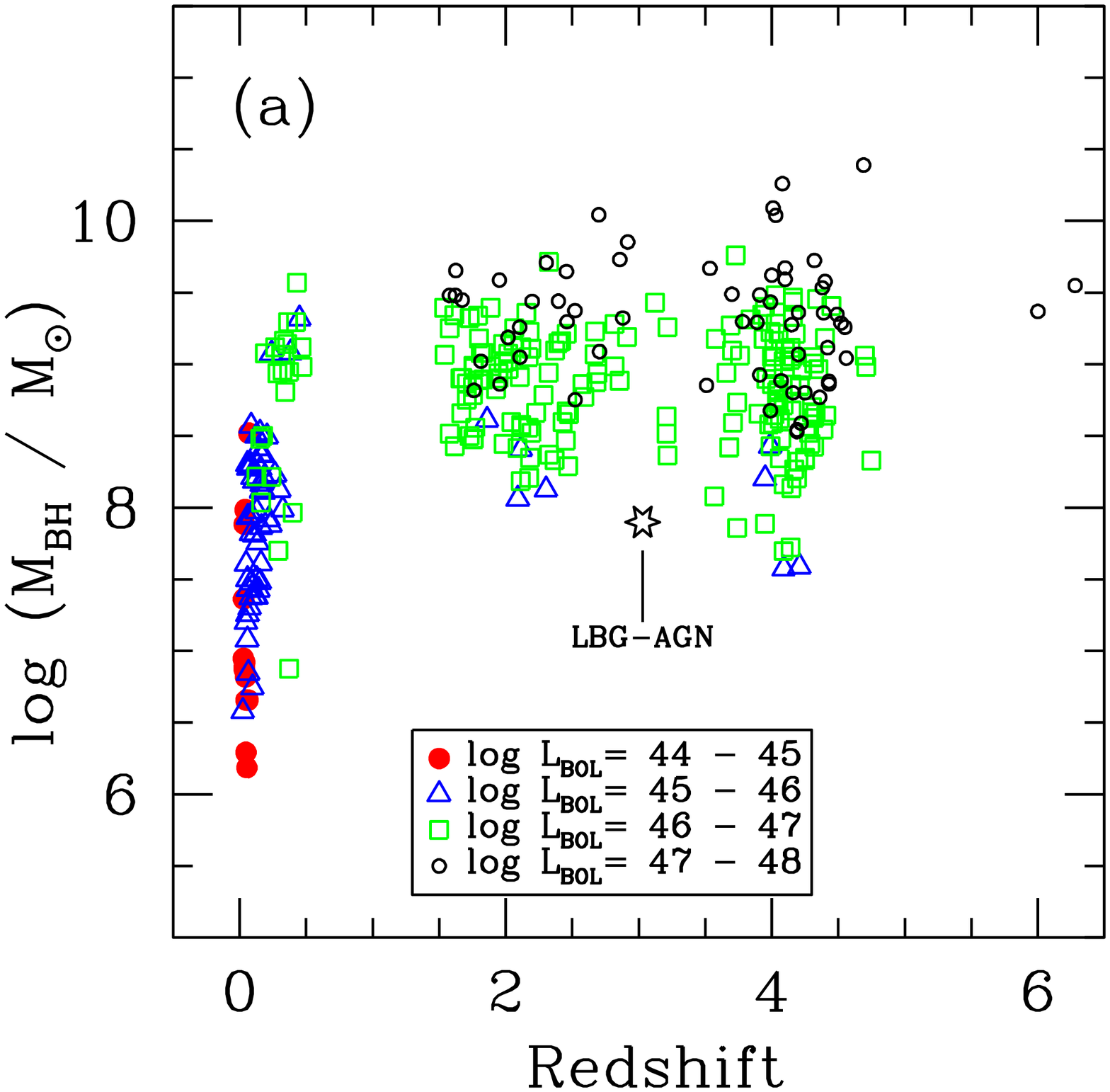} }
\vspace{-7.0cm}\hspace{7.0cm}
\hbox{
\epsfxsize=7.0cm
\epsfbox{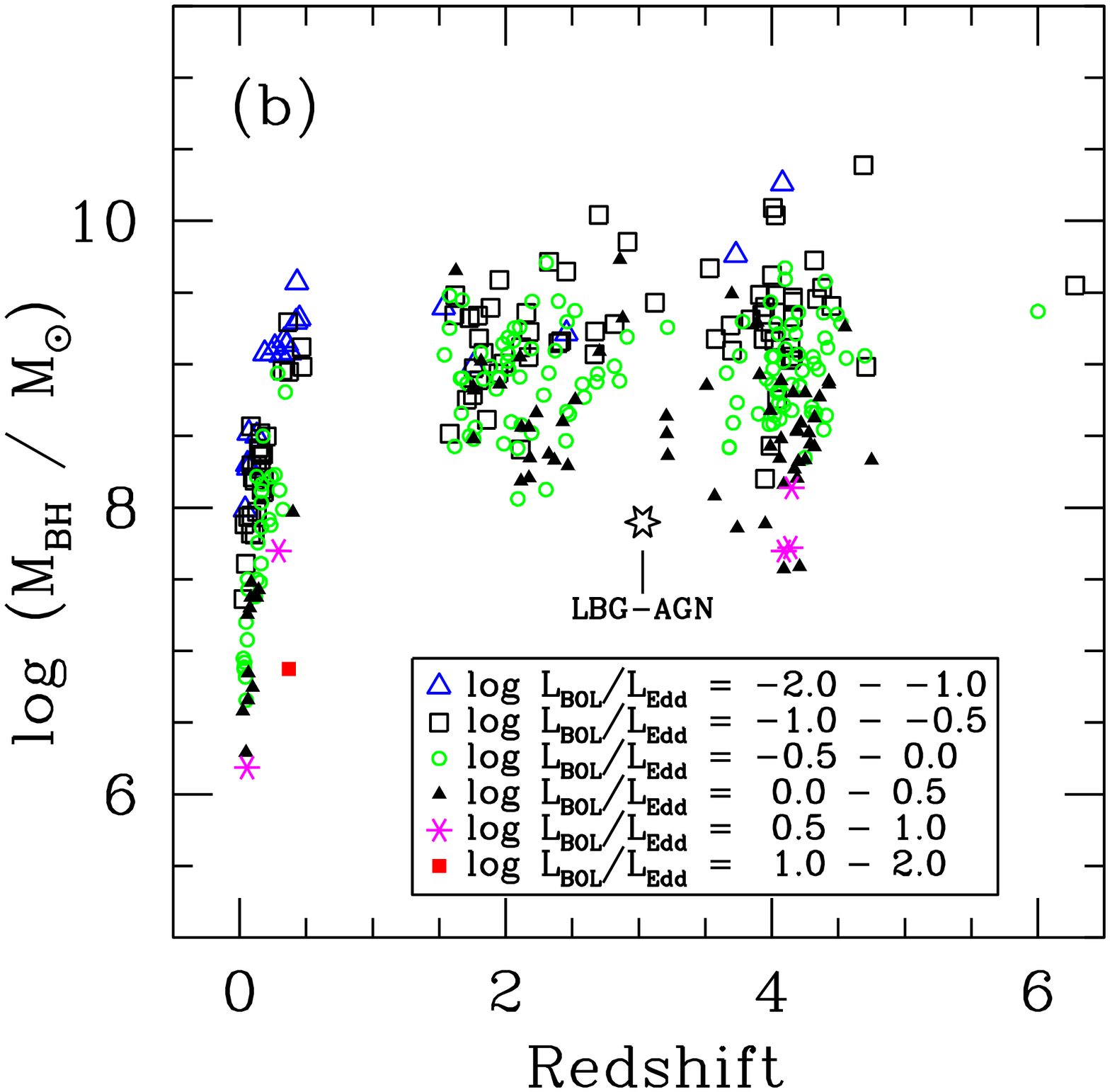}
}}
\caption[]{Distribution of \mbh{} with redshift shown binned in
({\it a}) \lbol{} and ({\it b}) \lol.  Error bars are omitted for clarity.
The labeled open star marks the likely position of an ``average'' broad-lined
Lyman-break galaxy AGN (see \S~\ref{qhosts}).
\label{m_zlbins.fig}}
\end{center}
\end{figure*}

\begin{figure*}[t]
\begin{center}
\vbox{
\hbox{
\hspace{1.3cm}
\epsfxsize=7.0cm
\epsfbox{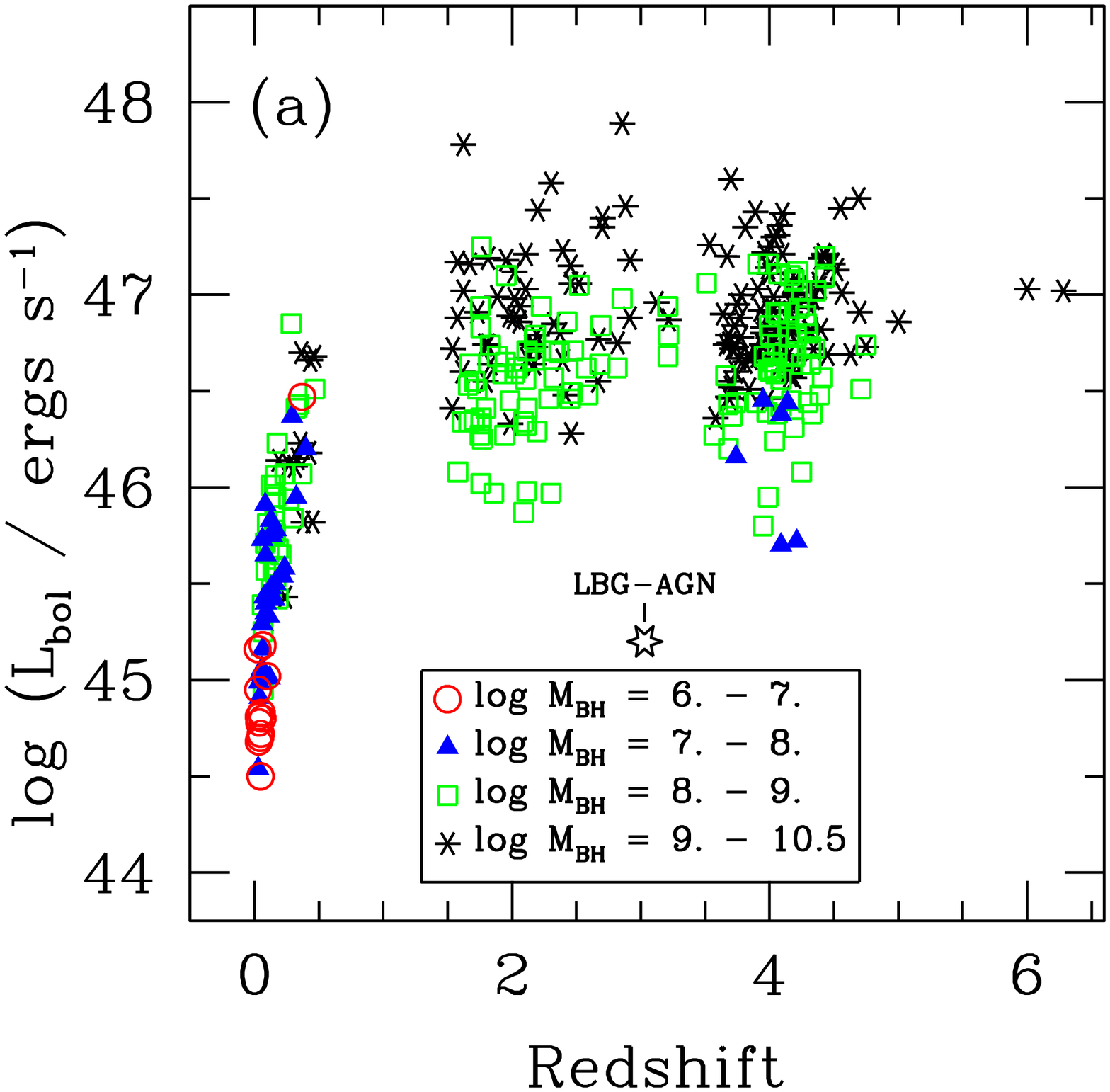} }
\vspace{-7.0cm}\hspace{7.0cm}
\hbox{
\epsfxsize=7.0cm
\epsfbox{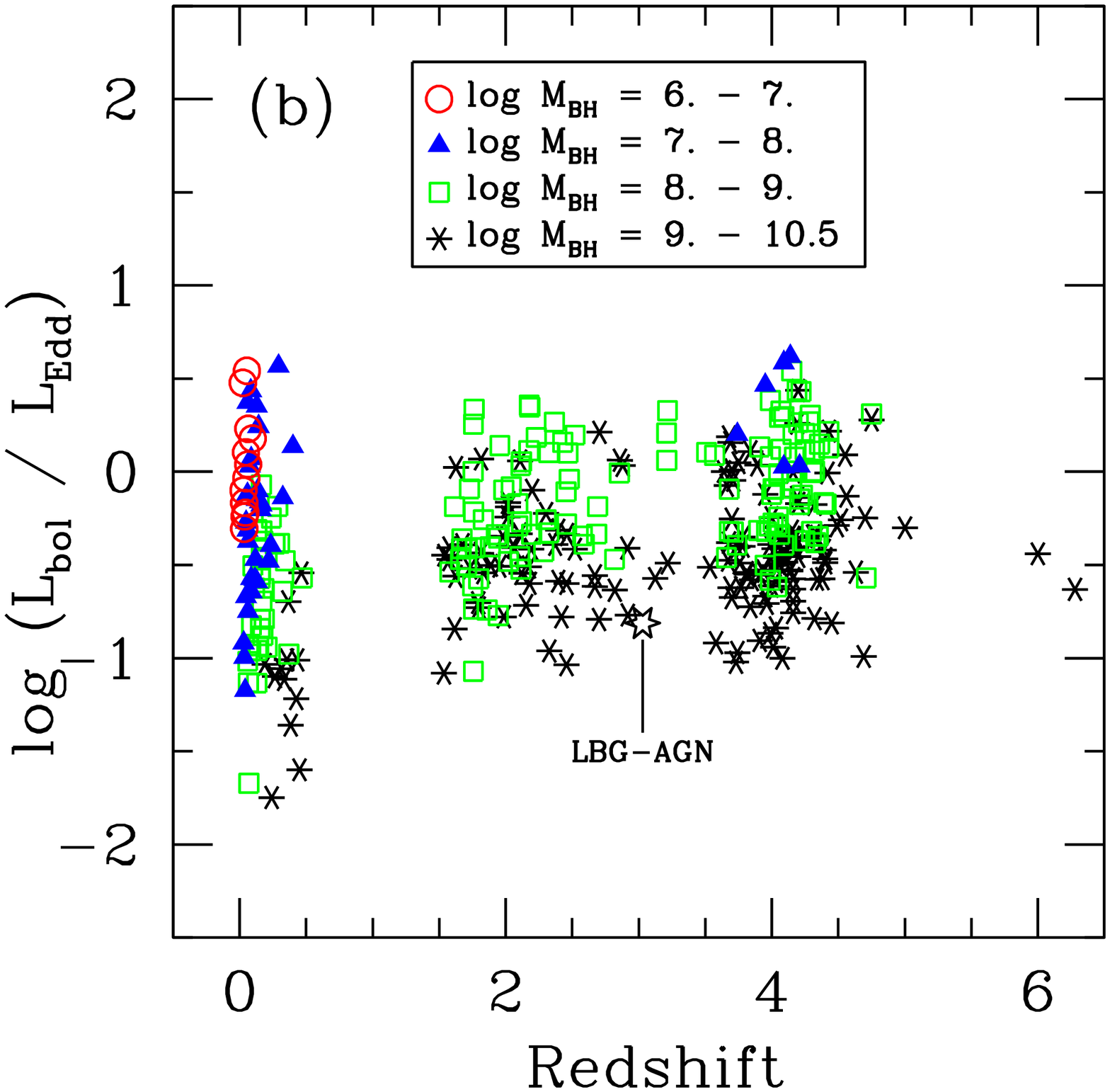}
}}
\caption[]{Distributions of ({\it a}) \lbol{} and ({\it b}) \lol{} with
redshift shown binned in \mbh.  Error bars are omitted for clarity.
The labeled open star marks the likely position of an ``average'' broad-lined
Lyman-break galaxy AGN (see \S~\ref{qhosts}).
\label{l_zmbins.fig}}
\end{center}
\end{figure*}

Figures~\ref{l_z.fig} and~\ref{m_z.fig} show how \lbol{} and \mbh{} 
distribute with redshift.  The curve segments are lower limits imposed 
by the survey limits or sample selection. 
In Figures~\ref{m_zlbins.fig} and~\ref{l_zmbins.fig} the \mbh, \lbol,
and \lol{} distributions with redshift are shown binned in different 
ways and with measurement uncertainties omitted for clarity.

The lower envelope of the observed \lbol{} values is easily explained 
by the survey or selection flux limits (Fig.~\ref{l_z.fig}).
The bolometric luminosity corresponding to 
\lstar{} and 10\,\lstar and their evolutionary tracks 
[\ie $L^{\ast}_{\rm bol}(L^{\ast}_B,z)$ and $L^{\ast}_{\rm bol}(10\,L^{\ast}_B,z)$] 
are shown for 
reference, where \lstar{} is the characteristic (break) luminosity of 
the quasar luminosity function, $\Phi(L_B,z)$.  For simplicity, the 
parameterization of Boyle \et (2000) is adopted here, namely:

\begin{equation}
\Phi(L_B,z) = \frac{\Phi(L^{\ast}_B)}{(L_B/L_B^{\ast})^{\alpha} + (L_B/L_B^{\ast})^{\beta}},
\label{lf.eq}
\end{equation}
where $L_{\rm B}$ is the $B$ broad-band luminosity, and $\alpha$ and $\beta$ are the 
slopes of the faint and bright end of the luminosity function, respectively.
The $L^{\ast}_{\rm bol}(z)$ values were computed as 
$10^{1.6} \times L^{\ast}_{\rm B}(z)$. 
The bolometric correction factor to $L_{\rm B}$ was determined from the updated 
quasar SED described in \S~\ref{mlmsmts}.  The $L^{\ast}_{\rm B}$ and its 
evolution\footnote{The luminosity function determined by Pei (1995) is
consistent with that of Boyle \et (2000) within the parameter uncertainties.} 
was determined for $\sim$6000 quasars by Boyle \et (2000) between redshifts zero 
and 2.3 from the LBQS and 2dF Quasar Redshift Survey data. 
Boyle \et extrapolate $L^{\ast}_{\rm B}(z)$ to $z > 2.3$ based on the fit to 
the data at lower redshifts.  
Figure~\ref{l_z.fig} shows that the $3.5 \lsim z \lsim 5$ quasars 
studied here are very luminous: most are well above several \lstar{}
(see e.g., Fig.~\ref{l_z.fig}b), assuming $L^{\ast}_{\rm bol}(L^{\ast}_B,z)$
can be trusted for short extrapolations (from $z \approx$ 2.3).  
The extrapolation to $z \approx 6$ is much more uncertain, but, evidently, 
the most distant quasars detected are unusually bright. 
Also, the BQS quasars are clearly among the most 
luminous quasars at low redshift, as is well known, while the $z \approx 2$ 
quasars are, on average, \lstar{} quasars or slightly more luminous. 
A preliminary analysis of the Large Bright Quasar Survey (LBQS) data from
Forster \et (2001) of quasars at $1.3 \lsim z \lsim 3.0$ suggests that the
brightest of these quasars may exceed \lbol{} $\approx 10^{47}$ \ergs{} 
and \mbh $\approx 10^{10}$\Msol{}. If indeed so, the relatively lower 
\lbol{} values of the intermediate-$z$ sample (see Fig.~\ref{l_z.fig}) are 
consistent with the fact that these quasars were intentionally selected not 
to be the brightest at a given $z$ (Vestergaard 2000).

One possible explanation for the high luminosities is gravitational 
lensing by foreground galaxies too faint to be detected at present.
The lensing fraction is strongly dependent on the slope ($\beta$) of the
bright end of the optical luminosity function 
at $z > 3$ (\eg Wyithe \& Loeb 2002).
Fan \et (2001a) determined the luminosity function for the well-defined 
``color-selected sample'' of bright quasars at $z \approx 4$ and find a 
slope of $\beta$ = 2.58, somewhat flatter than the slope, $\beta = 3.43$, 
of the bright end of the $z \lsim 2.3$ luminosity function (Boyle \et 2000). 
This shows that the $z \approx 4$ ``color-selected sample'' contains a larger 
fraction of the most luminous quasars, in accordance with the findings here.
More importantly, this bright end slope allows a handle on the lensing
fraction at high-$z$, assuming the ``color-selected sample'' is 
representative of bright $z \gsim 4$ quasars.
Wyithe \& Loeb (2002) find a relatively high probability ($P$) that the 
luminosity of $z \approx 6$ quasars is enhanced by lensing: 
7\% $\lsim P \lsim$ 30\% for $2.58 \leq \beta \leq 3.43$, respectively. 
The luminosities of the $z \approx 4$ quasars are not expected to be 
significantly lensed: 4\% $\lsim P \lsim$ 13\%. These estimates show
that very few of the high-$z$ quasars studied here are likely to be
gravitationally lensed.
Even if as many as 13\% of the most luminous $z \approx 4$ quasars
are lensed, the average \lbol{} of the remaining quasars 
is still well above the predicted $L^{\ast}_{\rm bol} (z \approx 4)$: 
$<$\lbol$> \approx 10^{46.8}$ \ergs{} = 9.4 $\times < L^{\ast}_{\rm bol} (z)>$.
Evidently, with the current survey flux limits, these high-$z$ quasar samples 
only probe the brightest tail of the quasars possibly present at $z \approx 4$. 
Future surveys need to go much deeper\footnote{
An extrapolation of $L_B^{\ast}$ shows that 1.0 (0.01) $L^{\ast}$ quasars are
expected to have broad-band magnitudes: $i^{\ast} \approx$25 (30) mag at 
$z \approx 5$ and $z^{\ast} \approx$ 29 (31.5) mag at $z \approx 6$.
} 
to properly test whether or not the $z > 3$ luminosity
function is significantly different than that at lower redshifts.


Figure~\ref{m_z.fig} shows that even the high-$z$ black-hole masses are 
quite large\footnote{
The mass estimates of the highest redshift ($z$=6.41) quasar SDSS J114816.64$+$525150.3
based on both \mgii{} (Willott \et 2003a) and \civ{} line widths (Barth \et 2003) are
in agreement (but note that individual mass estimates based on scaling relationships 
may be uncertain by a factor 10 (Paper~I); the factor 3 uncertainty is valid for 
statistical samples only).
} and, in particular, do not show a decrease relative to the 
masses of the intermediate-$z$ quasar sample. However, the masses of the 
$z \approx 2$ LBQS quasars may reach 10$^{10}$\Msol{}, as noted earlier. 
Thus, a slight mass decrease may exist at $z > 3.5$; the apparent upper 
envelope in the quasar masses at $4 \lsim z \lsim 5$ furthermore suggests this.
Also shown in Figure~\ref{m_z.fig} is
the relative position of \mstar($z$), 
determined from the evolution of $L^{\ast}_{\rm bol}$ assuming 
$L^{\ast}_{\rm bol} \equiv L^{\ast}_{\rm Edd} \propto$\,\mstar.  
Since it is unclear whether all quasars 
radiate at the Eddington luminosity, the relative location of \mstar{}($z$),
computed assuming $L^{\ast}_{\rm bol} \equiv 0.1 L^{\ast}_{\rm Edd}$, is 
also shown for reference.  If \lbol{} = $L^{\ast}_{\rm Edd}$ is valid, most of 
the quasars across the observed redshift range are very massive, reaching
\mbh{} $>$ 10 $-$ 15 \mstar. This is true even at the lowest redshifts 
($z < 0.6$).  Regardless of the true value of \mstar ($z \approx 4$), the 
$z > 3.6$ quasars stand out as being extremely massive.  For most quasars, 
\lbol{} = 0.1 $L^{\ast}_{\rm Edd}$ may be reasonable, given previous 
estimates at lower redshifts (\eg Wandel \et 1999; 
Awaki \et 2001; Czerny \et 2001; but see also Collin \et 2002).
In this case, the BQS masses distribute almost evenly about the predicted
value of \mstar{}, with a range of a factor $\sim$10 in each direction.
Also, the values of \mbh{} for the $z \approx 2$ quasars are mostly in the range
0.1 $-$ 1 $\times$ \mstar{}. Furthermore, if the \mstar($z$) extrapolation to 
$z \approx 4$ is reasonable, then the $z > 3.6$ quasars have masses in the range 
$\sim$ 0.1 $-$ 10 \mstar{}, similar to the BQS.  The average 
observed Eddington ratio is, however, well above 0.1 (Fig.~\ref{mlhistzbins.fig}) 
--- possibly closer to a value of 1 for the highest redshift quasars --- with a 
typical \lol{} value of 0.4 $-$ 0.5 at $z > 1.5$. Adopting this ratio would 
make \mstar{} fall about halfway between the two \mstar($z$) curves shown in 
Figure~\ref{m_z.fig}, making the $z \approx 2$ quasars \mstar{} on average and 
the $z > 3.6$ quasars as massive as 10 $-$ 15 \mstar{} or even larger.
Note that the absence of much less massive quasars, especially at high-$z$,
is mainly due to the selection and survey limits (Figure~\ref{m_z.fig}). 
Also, the high \mbh{} values are not due to severe overestimates
caused by the adopted method to estimate \mbh{} (\S~\ref{mreliability}). 


The very detection of the $z \gsim 3.6$ quasars means they are very
luminous, as already discussed. But, must they also necessarily be massive?
Rejecting the possibility that they are grossly super-Eddington, the
obvious possibilities for artificially enhancing their \mbh{} estimates
are (a) gravitational lensing, and (b) relativistic beaming. Gravitational
lensing can account for some, but certainly not all quasars (30\% at the
very most), as argued above. The high-$z$ quasars are also not likely
to be strongly beamed, judging by the similar appearance of their UV spectra
to those of low-$z$ AGNs: beamed sources (BL\,Lac and OVVs) have very weak or
absent emission lines (\eg Peterson 1997). The equivalent width ($EW$) of
the UV lines in the high-$z$ quasars are not much different from those of
low-$z$ AGNs. The average $EW$(\lya$+$\nv) and $EW$(\civ) in the
Constantin \et sample are 80\AA{} and $\sim$40\AA{}, respectively, while
Fan \et (2001b) find a mean $EW$(\lya$+$\nv) of 69\AA{} for the ``color-selected
sample''. These compare very well with the typical $EW$s in low-$z$ AGNs:  
$EW$(\lya$+$\nv) = 75\AA{} and $EW$(\civ) = 35\AA{} (Peterson 1997). 
In fact, the $EW$s at low and high-$z$ are remarkably similar considering 
more luminous quasars are expected to have lower line $EW$ (the Baldwin effect;
\eg Baldwin 1977). 
So to conclude, since the $z \gsim 3.6$ quasars are rather luminous, they
must be massive as well.
Additionally, the large emission-line widths (Figure~\ref{fw_lum.fig}) also 
indicate the black-hole masses are large.

Figure~\ref{m_zlbins.fig} shows that even lower-luminosity
($\sim 10^{44}$\ergs{}) quasars can have relatively high masses,
$\gsim 10^8$\Msol{} (the BQS at $z < 0.6$). 
The tendency for \lol{} to increase with decreasing \mbh{} is
particularly prominent for the $z < 0.6$ objects (Fig.~\ref{m_zlbins.fig}b 
and~\ref{l_zmbins.fig}b).  Notably, even the 
most massive $z > 1.5$ quasars also have high Eddington ratios 
(Fig.~\ref{l_zmbins.fig}b). Also, a large fraction of 
the \mbh $\approx 10^8 - 10^9$\Msol{} quasars appear to have super-Eddington 
luminosities (Figs.~\ref{m_zlbins.fig}b and~\ref{l_zmbins.fig}b).
For the distant ($z > 1.5$) quasars, there is considerable overlap in
luminosity for the two highest mass bins (Fig.~\ref{l_zmbins.fig}a)
and visa-versa (Fig.~\ref{m_zlbins.fig}a).

\subsection{Overview of Results \label{results}}

A number of conclusions can immediately be made based on the 
distributions described above.

(1) At low redshift ($z < 0.6$), the brightest quasars, represented by the BQS 
quasars, exhibit a clear upper envelope in \mbh{} (Fig.~\ref{m_z.fig}
and~\ref{m_zlbins.fig}) and \lbol{} (Fig.~\ref{l_z.fig} and~\ref{l_zmbins.fig})
which increases with redshift. Although such a sharp upper cutoff is not seen 
for the more distant quasars there does appear to be a real physical ceiling 
of \mbh{} $\approx 10^{10}$\Msol{} and \lbol{} $\approx 10^{48}$ \ergs{} at 
all redshifts.  The \lol{} ratio evenly covers the same approximate range 
($0.1 \lsim$ \lol{} $\lsim 3$) at all redshifts (Figs.~\ref{mlhistzbins.fig} 
and~\ref{l_zmbins.fig}).
The lower envelopes of \mbh, \lbol, and \lol{} for all the quasars 
are due to the original discovery surveys being flux limited. 

(2) The high-$z$ quasars have relatively high values of \mbh{} and \lbol.
The central masses of the highest redshift ($z > 3.6$) quasars do not
drop significantly compared to the intermediate redshift ($z \approx 2$)
objects. However, the $z \approx 2$ quasars were not selected to be 
the brightest, and therefore the most massive, at this redshift 
(Figs.~\ref{mlhistzbins.fig}, ~\ref{l_z.fig}, and~\ref{m_z.fig}). 
A preliminary analysis of the $z > 1.3$ LBQS sample suggests that a
mild decrease in, at least, \mbh{} may exist at $z > 3.6$, as the
brightest $z \approx 2$ LBQS quasars may exceed \mbh $\approx 10^{10}$\Msol{}
and \lbol{} $\approx 10^{47}$ \ergs.
It is particularly striking that the most distant quasars at $z \approx 6$
have similarly high \mbh{} $\approx 10^9$\Msol{} as at lower redshifts.
These quasars do have a high probability ($P \lsim$30\%) of being 
gravitationally lensed (Wyithe \& Loeb 2002). But if they are not lensed,
the known $z \approx 6$ quasars must be very massive.  It has been argued 
that such massive quasars must necessarily reside in massive dark matter
halos and be associated with very rare, high density peaks in the dark 
matter distribution at that epoch (\eg Efstathiou \& Rees 1988; 
Fan \et 2001c; Volonteri \et 2002).

(3) Most of the objects have masses above \mstar($z$), even at the 
lowest redshifts where $L^{\ast}(z)$ is reliable. 
Given the mean \lol{} at $z > 1.5$, it follows that the actual \mstar{}
in Fig.~\ref{m_z.fig} probably lies between the two dashed curves. 
This makes the quasar masses of the intermediate-$z$ sample on 
average \mstar{}
and makes most of the $z \gsim 3.6$ quasars 10 $-$ 15 \mstar{}, if the
luminosity function can be reasonably extrapolated to $z \approx 4$.

\section{Discussion
\label{bhprops}}

It was argued in section~\ref{mlmsmts} that black-hole masses of distant
AGNs and quasars can be predicted using their broad-line widths, their
luminosities, and scaling relationships. For a statistical sample of objects
this method appears accurate to within a factor of about 3. 
When this method is applied to the quasar samples described in        
\S~\ref{data} it is concluded that quasars at $4 \lsim z \lsim 6$ have
black-hole masses of $10^8 - 10^{10}$\Msol{}, with an average of about 
$10^9$\Msol{} (\S~\ref{distributions}). In spite the fact that the 
currently known high-$z$
quasars represent only the upper end of the luminosity function (\eg Fan 
\et 2001a) (and presumably the upper end of the mass function), it is
important to appreciate that black-hole masses as large as a few times 
$10^9$\Msol{} are capable of forming by $z = 6$. Certainly, the high
luminosities of these quasars suggest this must be the case (Fan \et 
2001c), but the present results provide evidence of high black-hole masses
independently of arguments based only on the Eddington limit.

It is currently unknown how these massive black holes form or on what 
time scale. As argued by Fan \et (2001c), it takes 20(\lol)$^{-1}$ 
$e$-folding times to grow a 10\Msol{} black hole with radiative efficiency 
of 100\% to a mass of $\sim 5 \times 10^9$\Msol{}. As this time scale 
is within, although close to, the age of the Universe at $z \approx 6$ for 
a cosmology of $H_0$ = 65 ${\rm km~ s^{-1} Mpc^{-1}}$, $\Omega_0$ = 0.35, 
and $\Lambda$ = 0.65, this growth scenario is not ruled out by its time
scale.

These conclusions depend on whether or not the scaling relations, defined for
nearby AGNs and quasars, can be applied to luminous and very distant quasars.
The main justifying reason is that the currently available data at radio, 
UV, and X-ray wavelengths show no clear evidence that the properties 
of $z \gsim 4$ quasars are any different than those of other luminous 
quasars known at lower redshift. First, quasar rest-frame UV spectra are very 
similar at all redshifts up to at least $z \approx 4.5$ (\eg Constantin \et 
2002; Dietrich \et 2002a; barring luminosity effects such as the Baldwin effect). 
Also, as discussed in \S~\ref{distributions}, the values and distributions
of \mbh{}, \lbol, and \lol{} are also similar, especially at $z > 1.5$.
Second, at radio energies, the SDSS $3.6 \lsim z < 5$ quasars exhibit 
similar properties ($\sim$10\% fractional occurrence and similar radio spectra) 
as known for $z < 3$ quasars (Carilli \et 2001a; Stern \et 2000; see 
Petric \et 2003 for corroborating results for $z > 5$ quasars). 
Moreover, the X-ray properties of high-$z$ quasars (\eg the optical-UV to 
X-ray slope, $\alpha_{\rm OX}$, and X-ray slope, $\Gamma_x$) are probably 
not much different from those of their lower-$z$ cousins. 
Brandt \et (2002) and Mathur, Wilkes, \& Ghosh (2002) independently
concluded this for the $\alpha_{\rm OX}$ values of three $z \approx 6$
quasars. Bechtold \et (2003), on the contrary, argue that
$z \gsim 4$ quasars have relatively steeper $\alpha_{\rm OX}$ values 
($\gsim 1.5$ with an average of $1.6\pm0.1$) based on \chandra{} data of 
17 optically selected, radio-quiet ($3.7 \leq z \leq 6.3$) quasars, but 
their $\alpha_{\rm OX}$ values are slightly overestimated\footnote{
Bechtold \et (2003) overestimate the $\alpha_{\rm OX}$ values by 
adopting a UV slope of $-$0.3. Fan \et (2001b) find that an average slope 
of $-$0.8 is more appropriate for $z \approx 4$ quasars.
For example, for two of the objects studied by Bechtold \et the UV slopes are
measured by Fan \et and the resulting $\alpha_{\rm OX}$ are 1.47 and 1.48,
much closer to the average of 1.43 measured at lower redshifts (Elvis \et
2002; note that the radio-quiet quasars observed with \rosat{} have
$<\alpha_{\rm OX}> = 1.55$; Yuan \et 1998). It may thus not be the norm
that high-redshift quasars have very steep optical-X-ray slopes; but see 
Vignali \et 2003).}. Also, the $\alpha_{\rm OX}$ slope extends to 
flatter values ($\alpha_{\rm OX} \gsim 1.2$) when $z \approx 4$ quasars 
from other studies are included (Fig.~1 by Bechtold \et), similar to average 
$\alpha_{\rm OX}$ values (1.55 $\lsim$ $<\alpha_{\rm OX}>$ $\lsim 1.65$) 
for $z < 3$ radio-quiet quasars (Yuan \et 1998).
However, Vignali \et (2003) also report steeper $\alpha_{\rm OX}$ values 
for their $z \approx 4$ quasars, but argue it is a luminosity effect. 
Nevertheless, Vignali \et find the X-ray and broad-band spectra similar
to those of low-$z$ quasars.
Bechtold \et also determine X-ray slopes of $\Gamma_x \approx 2.5$ (1.5) 
for their $4.0 \leq z \leq 4.5$ ($z \gsim 5$) quasars. These slopes are 
somewhat flatter than the {\it steepest} slopes of 3 to 4 observed for $z \lsim 2$
\rosat{} quasars. However, this apparent flattening of $\Gamma_x$ with
redshift is most likely a selection effect because different energy ranges
of the X-ray spectrum are observed at different redshift, as also suggested by 
Bechtold \et (2003).
In conclusion, the apparent similar properties across the SED of distant 
quasars ($z > 1.5$) indicate that the properties of the central engine of
the high-$z$ luminous quasars are sufficiently similar that the scaling 
relationships can used to estimate black-hole masses of these objects.
The efficacy of this method \mbox{is discussed in more detail in 
\S~\ref{mreliability}.}

The second key result presented in \S~\ref{distributions}, is the
clear presence of a ceiling to the \lbol{} and \mbh{} values at all
redshifts. From redshift zero to 0.6, the maximum values of all three 
parameters increase to a certain value and then level off at higher 
redshifts. The upper envelopes in \mbh{} and \lbol{} are genuine, since 
none of the sample selections impose upper limits in the source brightness.
Also, there are no apparent selection biases preventing a detection of 
very broad-lined, bright quasars, with the possible exception of the 
$z \approx 2$ quasars (\S~\ref{distributions}). 
The apparent lack of very luminous and very massive quasars may be
due to either or both of the following explanations. First, very luminous
(massive) quasars are expected to be very rare, given the steep 
decline of the upper end of the luminosity (mass) function.
Second, the above-mentioned ceilings may be due to some physical
limit to black hole growth. Possible causes for such a maximum 
sustainable mass and luminosity include: (a) limits in the dark matter 
gravitational potential, (b) limits in the fuel supply rate, (c)
transition of accretion mode or timescale, and (d) changing accretion 
disk (\ie emission) properties with black-hole growth (\eg Caditz, 
Petrosian, \& Wandel 1991; Yi 1996; Kauffmann \& Haehnelt 2000).

Two recent studies (Bechtold \et 2003; Netzer 2003) report quasar masses
at high-$z$ ($z \approx 4$ and $z \lsim 3$, respectively) reaching or
exceeding $10^{10}$\Msol{} in apparent conflict with the above conclusion. 
However, these \mbh{} values are not inconsistent with the current analysis
once differences in cosmology and statistical uncertainties are taken 
into account, as explained in the following.
Firstly, Netzer (2003) finds a maximum mass of 10.4\,dex assuming an
$R - L$ slope of 0.7 and  $H_0$ = 70 ${\rm km~ s^{-1} Mpc^{-1}}$,
$\Omega_m = 0.3$, and $\Omega_{\Lambda} = 0.7$. For this cosmology, 
his \mbh{} values are about a factor 2 (or $\lsim 0.35$\,dex) higher
than for the cosmology adopted here 
($H_0$ = 75 ${\rm km~ s^{-1} Mpc^{-1}}$, q$_0$ = 0.5, and $\Lambda$ = 0).
Combining this with the probable uncertainty in his mass values of 
$\sim$0.6\,dex, given a measurement error of $\lsim 0.25$\,dex and a
statistical uncertainty in the method of $\sim$0.5\,dex (Paper~I),
the highest mass values of Netzer's sample and those analyzed here are
statistically consistent. An important point to keep in mind is 
that {\it individual} \mbh{} values can be in error by as much as
a factor 10. The fraction ($\lsim 9$\%) of quasars 
in Netzer's sample with \mbh{} $> 10^{10}$\Msol{} is also within the 
expected fraction ($\sim$10\%) of objects for which the \mbh{} values 
may be off by 1\,dex or more (Paper~I).
Secondly, the maximum \mbh{} value ($\sim$10.7\,dex) reported by 
Bechtold \et (2003) for 
$H_0$ = 50 ${\rm km~ s^{-1} Mpc^{-1}}$, q$_0$ = 0.5, and $\Lambda$ = 0,
corresponds to $\sim$10.45\,dex in the current cosmology. Although,
Bechtold \et do not quote their errors, the uncertainties in their \mbh{}
values are presumably as large as or even larger\footnote{This is owing to 
their crude correction of their FWHM(\civ) values to FWHM(\hb), so to apply 
the optical scaling relationship of Kaspi \et (2000) to obtain central 
masses, and their adoption of rather flat UV slopes ($-$0.3) relative 
to the steep slopes ($-0.8$) measured by Fan \et (2001b) for their 
representative $z \approx 4$ quasars (see footnote above).} than those 
of Netzer (2003), and, hence, their maximum \mbh{} value is similarly 
consistent with the upper boundary $\leq10^{10}$\Msol, seen in the 
current study, to within the errors.

Netzer (2003) also claims that \mbh{} $\gsim 10^{10}$\Msol{} is 
inconsistent with the locally defined \mbh{} $- \sigma$ relationship
($\sigma$ is the stellar velocity dispersion in the host galaxy bulge),
since the implied $\sigma \approx 700$\kms{} is not observed locally.
As Netzer notes, either (1) the \mbh{} values are overestimated, because,
for example, the $R - L$ relationship does not extrapolate to high-$z$
quasars, or (2) the \mbh{} $- \sigma$ relationship changes with redshift.
It is argued in \S~\ref{mreliability} that the $R - L$ relationship is
not likely to be misleading and that the scaling relationships used 
here yield reasonable \mbh{} estimates to within a factor $\lsim$4.
Furthermore, the lack of local black-holes with masses $\sim 10^{10}$\Msol{}
does not mean that the large \mbh{} values inferred from the scaling
relationships are strongly overestimated. 
The reason is that in addition to a possible evolution of the \mbh{} $- \sigma$ 
relationship and the statistical uncertainty in the \mbh{} estimate,
the lack of local black-holes with masses $\sim 10^{10}$\Msol{} 
can possibly be explained by 
the relatively small volume probed in the local Universe: for the 
comoving space density of luminous, massive ($\gsim 5 \times 10^9$\Msol)
quasars at $z \approx 4$ of $\lsim 12.5$ Gpc$^{-3}$ (Vestergaard \& Osmer,
in preparation), a volume approximately 20 times larger than that currently
probed locally ($\sim$100\,Mpc) is needed to detect one such massive
quiescent black-hole. 

\subsection{Reliability of the Mass Estimates for High Redshift Quasars 
\label{mreliability}}

It is a serious concern whether the scaling relationship, defined 
for a representative sample of nearby AGNs and quasars, can rightfully 
be applied to luminous, distant quasars. Specifically, can extension 
of this relationship cause the masses of the distant quasars 
\mbox{($\sim 10^9$\Msol{})} to be severely overestimated?
Netzer (2003) also points out that the high masses he finds at $z \lsim 3$ 
appear inconsistent with the local \mbh{} $- \sigma$ relationship. 
In this section, the efficacy of the scaling relationships for luminous, 
distant quasars is discussed.  It is argued that application of the 
scaling relationships to such quasars is valid, owing mainly to the 
similarity of quasar spectra across the known redshift and luminosity 
ranges.  
In particular, the scaling relation is unlikely to be 
inaccurate by a factor $\gsim$4 in the mass.

The mass estimates will obviously be misleading if the assumptions
made when extending the scaling relationships beyond the nearby AGNs
break down.  The two assumptions for 
\mbh{} $\propto$ FWHM(\civ)$^2$ $\lambda L_{\lambda}^{0.7}$(1350\AA)
are: (1) that the size, $R_{\rm BLR}$, of the
(\civ) emitting region in the broad line region -- \ie its distance 
from the continuum source -- can be estimated reliably from the continuum
luminosity, $L_{\lambda}$(1350\AA), and (2) that the line width is a
reasonable proxy for the gravity-dominated velocity dispersion of the
line-emitting gas. The critical issues are therefore: 
(a) whether or not the $R - L$ relationship is valid at higher redshift 
and for more luminous quasars than the sample for which it was defined, and 
(b) whether or not the \civ{} emission gas is virialized or, for example, 
the \civ{} profile has strong blue asymmetries which may be a sign of BLR 
outflows as suggested for the narrow-line Seyfert~1 (NLS1) galaxies
(\eg Leighly 2000). Each issue is addressed in turn below.

First, the $R - L$ relationship is not likely to be much different
for the more luminous and more distant AGNs. The reasons are twofold:
(1) the $R - L$ relationship is not extrapolated very far in luminosity:
for the highest $\lambda L_{\lambda}$(1350\AA) values of 47.4\,dex,
the $R - L$ relationship is extrapolated by less than 1.5\,dex, a
small fraction of the luminosity range of $\gsim$4\,dex, over which 
the $R - L$ relationship is defined 
($10^{42} \lsim \lambda L_{\lambda}$(5100\AA)/\ergs $\lsim 10^{46}$),
and (2) AGN and quasar spectra look very similar at all redshifts and 
luminosities considered here, \ie line flux ratios, line equivalent
widths, and SEDs are all similar (\eg Elvis \et 1994; Kuhn \et 2001;
Constantin \et 2002; Dietrich \et 2002a, 2002b; see also \S~\ref{mlmsmts}).
Specifically, if the $R - L$ relationship does
{\it not} apply to luminous, distant quasars, the spectra of these
objects would look somewhat different than they do. 
This can be quantified by photoionization modeling.
In essence, if the scaling relationship (eq.~\ref{logmuv.eq}) results in 
a mass overestimate of, for example, a factor 10 for a quasar with \lbol{} 
$\approx 10^{47}$\ergs{},
the virial theorem shows that the \lya{} and \civ{} emission lines are
emitted so close ($\lsim$33\,light-days) to the ionizing source that
these lines are produced very inefficiently, and, in particular, 
\ciii{} emission cannot be generated.  But, the presence of \ciii{} in all
spectra of quasars (at least up to $z \approx 4.3$; Dietrich \et 2002a)
confirms the existence of low-density BLR gas subject to a lower ionizing 
flux from which \lya{} and \civ{} are also much more efficiently 
emitted\footnote{These conclusions are based on the line emission resulting 
from gas with a range of column densities subject to the ionizing flux 
from a source with \lbol{} $\approx 10^{47}$\ergs{} as quantified by the 
photoionization grids of Korista \et (1997).}:
this gas will necessarily be located at a greater distance from the
ionizing source, close to the distance inferred from the luminosity.

Second, there is no evidence that the kinematics of the BLR in high-$z$
AGNs are {\it not} dominated by gravity. 
This conclusion is based on the following 3 arguments.
\begin{enumerate}  
\item
Most, if not all, of the BLR gas\footnote{This has so far been
established for the following emission lines: \siiv \,\lam 1400, \civ 
\,\lam 1549, \heii \,\lam 1640, \ciii \,\lam 1908, \hb \,\lam 4861, 
\heii \,\lam 4686 (see \eg Peterson \& Wandel 1999).} 
is in virial motion around the 
central source, including the high-ionization UV lines: all the AGNs 
with suitable multiple emission line measurements demonstrate a robust
virial relationship between their line widths and variability time lags 
(Peterson \& Wandel 1999, 2000; Onken \& Peterson 2002); 
while only four AGNs are testable so far, the general similarity 
of AGN spectra (\ie line shape, line widths, and variability properties)
argues that this relationship extends to most AGNs. 
\item
The distribution of \civ{} line widths in spectra of distant, luminous 
quasars is similar to that for nearby AGNs (typical values $\sim$4500\kms{}
with a range from 2000 $-$ 10000\kms{} in FWHM), including those for which 
the scaling relations are calibrated. Also, the \civ{} widths of the nearby 
BQS quasars are consistent with their \hb{} line widths, assuming virial 
BLR kinematics and that \hb{} is emitted twice as far from the central 
source as \civ{}, as established for three of the four AGNs discussed in 
item 1 (see also Korista \et 1995; see comment below on the one exception).
Furthermore, the similar FWHM(\civ)
distributions, where most are narrower than $\sim$8000\,\kms{} 
(\eg Figure~\ref{fw_lum.fig}), argue that the unknown, but possible,
contribution to single-epoch line widths from non-variable optically
thin emission (Shields, Ferland, \& Peterson 1995; Peterson 1997) is 
highly unlikely to make FWHM(\civ) in error by a factor $\gsim$2; 
such an error translates to a mass uncertainty of a factor $\lsim$4,
comparable to the statistical uncertainty (\S~\ref{mlmsmts}).
\item
Outflow of the BLR line-emitting gas is not important, at least
for the quasars considered here. This is supported by two facts:
(i) none of the \civ{} profiles of the $z > 1.5$ quasars 
(\S~\ref{thesisdata}, \S~\ref{hizdata}), and most quasars in general, 
clearly\footnote{Some spectra are noisy and the reality of such a 
prominent triangular line shape is difficult to discern.} 
resemble the broad, triangular, blue asymmetric \civ{} profiles of the
high luminosity NLS1s; those \civ{} profiles have been argued to signify 
strongly outflowing high-ionization line-emitting gas (\eg Leighly 2000), 
as NLS1s are very luminous for their \mbh, and
(ii) while a small fraction ($\lsim$20\%) of the \civ{} profiles in
the high-$z$ sample (\S~\ref{hizdata}) show weak, blue-ward asymmetries
(mostly suspected due to broad, blended \niv \lam 1486), those quasars
do not have much larger masses and luminosities than the remaining
quasars: a conservative estimate is an 0.2\,dex increase on average,
dominated by the lower-quality spectra of the Anderson \et sample;
the `asymmetric subset' of the Constantin \et and Fan \et quasars
exhibit average deviations of $\pm$\,0.1\,dex. 
Also, the \civ{} profiles of the $z \approx 2$ quasars are not 
typically asymmetric, and any asymmetries are restricted to
the very base of the profile and do not affect FWHM. Moreover, 
the strongest asymmetries do not extend blueward (Vestergaard 
2000; Vestergaard, Wilkes, \& Barthel 2000).
\end{enumerate}
In conclusion, blue asymmetries and UV outflows are generally not a 
concern for the \mbh{} estimates of quasars. Nonetheless, extreme care 
should be exercised when using the \civ{} line width to estimate the mass 
when signs of significant outflows are present. For this reason, mass 
estimates based on \civ{} are not well suited for the most luminous NLS1s.

It is important to emphasize that although all AGNs so far tested
show the BLR is virialized (\ie the virial product $R v^2$ is
constant for all emission lines measured), for one object (3C\,390.3) 
\civ{} is narrower than \hb{} and has a larger lag in contrast to
the other three objects. This behavior is not understood, but 
3C\,390.3 may have an unusual BLR structure as it is the only 
broad-line radio-galaxy with monitoring data and it, moreover, has 
double-peaked Balmer lines (Peterson \& Wandel 2000). Nevertheless, 
it is very important for the general applicability of the scaling 
relations that future studies address this issue.

\subsection{Do Black Holes ``Mature'' Before Their Host Galaxies Do? 
\label{qhosts}}

Given that black holes with large masses appear as early as $z \approx 6$ and 
given the intimate connection between the formation of galaxy bulges and their 
central, supermassive black holes inferred from the \mbh{} $- \sigma$ 
relationship, the question naturally arises whether or not this close 
connection also exists at high redshift.
Specifically, do these ``mature'' black holes (\ie those that have reached, 
say, $\sim 10^8 - 10^9$\Msol) also
reside in ``mature'' host galaxies (\ie virialized systems with the majority 
of their stars already formed) with aged ($>$1\,Gyr) stellar populations?
Among the currently available data on high-$z$ AGN host galaxies, there is 
considerable circumstantial evidence that massive active black 
holes may, in fact, reside in galaxies that are {\it not} fully relaxed, 
virialized systems and where many of the stars are yet to be formed.
This is based on
(a) the inferred high star-formation rates (SFRs) from the existing large masses 
of cool dust, (b) the presence of significant amounts of molecular gas 
in some sources, (c) the presence of large-scale young starbursts in at
least one well-studied $z \approx 4$ radio-galaxy with a large dust mass, 
(d) the strikingly different morphology of host galaxies of high-$z$ 
radio-quiet quasars and radio galaxies compared to their lower-$z$ cousins, 
and (e) the presence of massive, active black-holes in small, young galaxies 
at $z \gsim 3$.  Each issue is discussed in turn below.

\subsubsection{The Evidence}
\paragraph{Massive Star Formation in High-\mbox{\boldmath{$z$}} Quasars}

High-$z$ AGN host galaxies are apparently experiencing massive, large-scale 
star formation: the far-IR/sub-mm SEDs of $z \approx 4$ quasars 
(\eg Benford \et 1999; Carilli, Menten, \& Yun 1999; Carilli \et 2001a) are similar to 
those of ultra-luminous IR galaxies (ULIRGs) and starburst galaxies 
(\eg Downes \& Solomon 1998; Genzel \et 1998; Rowan-Robinson 2000; Klaas \et 2001). 
ULIRGs are dominated by extreme rates of star formation 
($\gsim$1000\,\Msol{} yr$^{-1}$) that yield unique far-IR/sub-mm SEDs, 
which are independent of whether or not a central AGN 
is present (\eg Downes \& Solomon 1998; Klaas \et 2001).
The rather cool ($T$\,$<$100\,K; typically $\sim40 - 50$\,K) 
starburst-heated dust dominates the far-IR emission (\eg Rowan-Robinson 2000; 
Dunne \et 2000; Klaas \et 2001), but AGN-heated dust 
($T$\,$>$100\,K) emits in the near-IR, shortward of $\sim$$60\mu$m                                                  
(\eg Wilkes \et 1999b; van Bemmel \& Dullemond 2003; Figure~8a by Farrah \et 2003).
Large amounts ($\gsim 10^8$\Msol) of cool ($\sim40 - 50$\,K) 
dust are commonly inferred\footnote{The inferred dust masses are, 
in fact, lower limits as cold dust ($T < $30\,K) is a poor emitter 
(\eg Omont \et 2001; Chapman \et 2003); there is probably a factor 
$\sim$3 more gas (\eg Dunne \et 2000).} from the restframe 
far-IR/sub-mm emission from $z \approx 4$ quasars (\eg Benford \et 1999; 
Carilli \et 2001a; Omont \et 2001; Priddey \& McMahon 2001; Isaak \et 2002). 
Large dust masses are often taken to imply high
SFRs, $\sim$1000 $-$ 2000 \Msol{} yr$^{-1}$ 
(Benford \et 1999; Carilli \et 1999, 2000, 2001b, 2003); 
such high SFRs are even inferred for the undetected sources (Isaak \et 2002).
Also, non-thermal radio emission contributes little to the shape of 
the cm to sub-mm SED, confirming its origin in massive star formation
(Carilli \et 2000, 2001a, 2001b; Yun \et 2000).

\begin{figure*}[t]
\begin{center}
\vbox{
\hbox{
\epsfxsize=8.5cm
\epsfbox{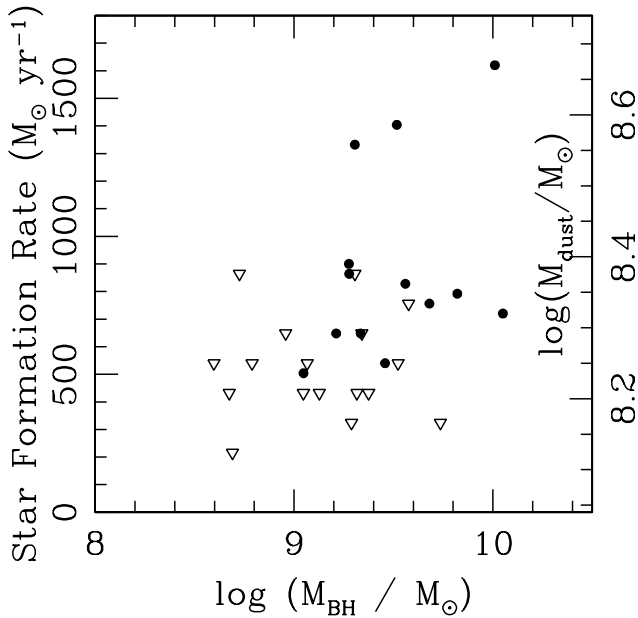}
\hspace{1.25cm}
}\vspace{-6.5cm}\hspace{9cm}\hbox{
\begin{minipage}{8cm}
\caption[]{Distribution of inferred star formation rate (SFR) and mass of dust
with estimated black-hole mass for a subset of the $z \approx 4$ SDSS quasars
analyzed here.  The far-IR emission emitted by the cool dust was observed by
Carilli \et (2001a).
Detections at $\lambda_{\rm obs} =$ 1.2\,mm are shown
as solid circles, while the open triangles denote $3\,\sigma$ upper limits.
The dust masses and SFRs are here computed using the prescription of Omont \et
(2001). Specifically, SFR $\approx 360\, \delta_{\rm MF}\, \delta_{\rm SF}
\,(S_{250\,\rm GHz}/mJy)$ \Msol{} yr$^{-1}$, where $\delta_{\rm SF}$ is the
ratio between the starburst's total luminosity and the far-IR luminosity, and
$\delta_{\rm MF}$ is ``a function of the present mass composition of the
stellar population'' (see Omont \et for explanation) and has values in the
range $0.8 \lsim \delta_{\rm MF} \lsim 2$.
Since starburst heating dominates the far-IR, $\delta_{\rm SF} \gsim$ 70\% is
assumed, and thus $0.6 \lsim \delta_{\rm MF} \delta_{\rm SF} \lsim 1.4$;
a fair estimate is $\delta_{\rm MF} \delta_{\rm SF} \approx 1$ yielding a
{\it lower limit} on the SFR.
\label{sfr.fig}}
\end{minipage}
}}\end{center}
\vspace{0.7cm}
\end{figure*}

The far-IR properties of the luminous $z \approx 4$ SDSS quasars
studied here appear very similar to those of other $z \approx 4$ 
quasars with far-IR data and to far-IR selected AGNs. 
Most of them have radio to far-IR SEDs consistent with 
starburst heating of large dust masses (Carilli \et 2001a), and
the inferred dust masses ($M_{\rm dust}$) and SFRs are similar to other 
$z \approx 4$ quasars:
Figure~\ref{sfr.fig} shows that for the subset of 30 quasars with 
\mbh{} and \lbol{} estimates, the 13 quasars detected at 1.2\,mm all 
have \mbh{} $\gsim 10^9$\Msol{}. Even for the undetected quasars, the 
upper limits cannot rule out the existence of SFRs of order
300 $-$ 700 \Msol{} yr$^{-1}$.

This line of argument is not without caveats.
For example, it can be argued (\eg Sanders \et 1989; Kuraszkiewicz \et 2003) 
that the IR SEDs of nearby quasars can be explained purely by AGN heating,
although this explanation is not universally accepted (\eg Farrah \et 2003;
Klaas \et 2001).
Nevertheless, no correlation is observed between quasar optical-UV
luminosity and far-IR luminosity,
as otherwise expected if AGN heating dominates the cooler dust  
(\eg McMahon \et 1999; Omont \et 2001; Isaak \et 2002; Priddey \et 2003).  
Also, Archibald \et (2002) argue that a high dust mass may indicate a
galaxy is in a late starforming phase ($\gsim$75\% complete),
as discussed below.

\paragraph{Molecular Gas.}

Huge amounts of cold molecular gas ($>10^{11}$\Msol{})  
are present in $z > 3$ far-IR luminous sources. This is 
(a) determined from CO emission line measurements 
(Ohta \et 1996; Omont \et 1996; Guilloteau \et 1999; 
Carilli \et 1999, 2002a, 2002b; Papadopoulos \et 2000), 
and (b) inferred from the dust masses and typical 
dust-to-gas ratios obtained from low-$z$ starbursts 
(\eg Sanders, Scoville, \& Soifer 1991; Dunne \et 2000).
Such vast gas reservoirs show that large amounts of stars are yet 
to be formed (\eg Solomon \et 1997; Downes \& Solomon 1998),   
and are evidence for ongoing massive 
star formation, especially when combined with large masses of cool 
dust (\eg Ohta \et 1996; Ivison \et 1998; Frayer \et 1999; 
Papadopoulos \et 2000; Carilli \et 2002a, 2002c).
In some cases, the cold gas clearly constitutes a significant fraction of 
the dynamical mass in the system, emphasizing the rather young nature of 
the galaxy  (\eg Ohta \et 1996; Guilloteau \et 1999; Solomon \et 1997; 
Downes \& Solomon 1998; Tacconi \et 1999).
For some high-$z$ sources, the sub-mm emission is 
(a) clearly separated (tens of kpc) from the AGN nucleus, and/or is
(b) distinctly extended 
(\eg Papadopoulos \et 2000; Omont \et 2001; Isaak \et 2002; 
Carilli \et 2001a, 2002b, 2003).
This is strongly suggestive of galactic-scale starforming regions and 
renders an AGN origin less likely (Carilli \et 2003).
Furthermore, when CO observations of high-$z$ sources are conclusive, 
the high masses of cold gas inferred from dust mass estimates are
confirmed (Hughes \et 1993; Ohta et al.\ 1996; Omont \et 1996; 
Papadopoulos \et 2000). 
Non-detections of CO line emission, however, do not necessarily rule 
out a young starburst interpretation because the observations are still
not very sensitive\footnote{The success-rate of CO detection is currently 
relatively 
low due to sensitivity limits of the detectors and the relatively 
narrow accessible band widths compared to the sometimes broad line 
widths (Ohta et al.\ 1996; Papadopoulos \et 2000; Omont 2003).          
Also, if the molecular gas is too cold, the CO line transitions are 
very weak.}.

\paragraph{Young Stellar Populations.}

For one far-IR/sub-mm luminous high-$z$ AGN, 
the $z \approx 3.8$ radio galaxy 4C\,41.17, stellar populations synthesis 
modeling has established the presence of large-scale young starbursts
with ages in the range 0.01\,Gyr to 0.4\,Gyr with a most probable 
age of 0.07\,Gyr (Dey \et 1997; Dey 1999). This is also important because 
it confirms the existence of starburst activity, expected based on the 
strong far-IR emission alone.
Although it is difficult to generalize based on one object, it is 
pertinent to emphasize that such analysis is very difficult to perform
for most high-$z$ AGNs, given the strong nuclear glare in quasars 
and the faintness of galaxies at high-$z$.   
Nevertheless, it is important for future studies to establish whether 
or not high-$z$ far-IR luminous AGNs in general have young starburst 
activity.

Archibald \et (2002) infer starburst ages of 0.5 $-$ 0.8\,Gyr for 4C\,41.17 
from a monochromatic luminosity ratio and their models of dust production 
and depletion with time in a young galaxy, and conclude the galaxy
is in a late starforming phase.  These ages are clearly at odds with the 
results of the more sophisticated age determination discussed above. 
Further developments in this area seem necessary to resolve these apparent 
inconsistencies.  More importantly, the measured
dust mass of 4C\,41.17 is {\it also} consistent with a starburst age in
the range 0.075\,Gyr to 0.095\,Gyr (Fig.~1 by Archibald \et 2002), in
good agreement with the results of Dey (1999). In conclusion, while dust
contents alone is an inaccurate measure of starburst ages, large dust
masses are not uniquely tied to $\gsim$1\,Gyr old galaxies but also 
appear in galaxies with starbursts as young as 0.04\,Gyr
(for $M_{\rm dust} \gsim 10^8$\Msol{}; Fig.~1 by Archibald \et 2002).

It is therefore an intriguing possibility that some high-$z$ AGN host
galaxies may have rather young stellar populations. However, a comparison
with the restframe IR emission is necessary to determine the extent to which
these young stars dominate the total stellar mass in place at that epoch.
The observed apparent extension of the $K - z$ relation, defined for
$z < 1.6$ radio galaxies (Lily \& Longair 1984), to $z \gsim 3$ radio
galaxies and quasars is sometimes used to argue for somewhat old stellar 
populations in their host galaxies. However, it is important to appreciate
that the increasing scatter in the observed relationship at $z > 2$ is 
{\it also} consistent with galaxies comprising mainly young stellar populations 
forming at $4 \lsim z \lsim 5$ (De Breuck \et 2002): the stellar mass in the host
galaxies (quantified by the $K$-band magnitude) is too low in many cases
to place all high-$z$ radio-galaxies on the $K - z$ relation with 
insignificant scatter. 
Also, it has been argued that the nature of proto-galaxies 
or young, less massive galaxies may also conspire to fall along the 
$K - z$ relation at $z \gsim 3$ (Eales \& Rawlings 1993; Willott \et 2003b).
Hence, it seems at least plausible that a good fraction of high-$z$ host
galaxies may be relatively young, even if this interpretation is not
universally accepted.

\paragraph{Host Galaxy Morphology}

While quasar host galaxies appear fully assembled by $z \approx 1$,
AGN host galaxy properties become increasingly more difficult to 
quantify at $z > 1$ (Kukula et al. 2001). Practically, radio galaxies are 
the only AGNs we can currently study beyond $z \approx 3$, owing to the
lack of nuclear glare that affects optical observations of quasar hosts. 
Radio-galaxy hosts do not appear fully assembled and dynamically relaxed.  
Specifically, in contrast to their lower-$z$ cousins with mostly single, 
smooth components, $z > 3$ radio galaxies display very clumpy structure: 
multiple, small star-forming components are embedded   
in large-scale ($\sim$50\,kpc) faint emission, indicative of galactic 
systems in early stages of assembly (\eg Graham \et 1994; van Breugel 
\et 1998; Pentericci \et 1999, 2001; Papadopoulos \et 2000).
Optically obscured $z > 2$ radio galaxies also
appear to be in an early stage of formation (Reuland \et 2003).

For high-$z$ radio-quiet quasar hosts, 
the best-quality imaging data available are likely those from NICMOS \hst{} 
studies of $2 \lsim z \lsim 3$ sources (Ridgway \et 2001; Kukula \et 2001). 
In contrast to radio-galaxy hosts, these are relatively structureless, 
single components. 
Also, at $2 \lsim z \lsim 3$, radio-quiet hosts are often more compact 
(scale-lengths $\lsim$4\,kpc and $\sim$2.3\,kpc on average; Ridgway et al.\ 2001) 
than (a) radio-loud hosts at that epoch (scale-lengths $\sim$ 11 $-$ 18\,kpc; 
e.g., Best et al.\ 1998; Kukula et al.\ 2001; Ridgway et al.\ 2001), and than 
(b) quasar hosts at $z \approx 0.2$ (mean scale-lengths of $\sim$8.2\,kpc; 
Bahcall et al. 1997; McLure et al. 1999). 
Moreover, the hosts of $z \approx 2$ radio-quiet quasars are apparently rather
young: Ridgway \et find the $2 \lsim z \lsim 3$ ($M_B \approx -24$ mag) radio-quiet 
quasar hosts contain only $\sim$10\% $-$ 20\% of their final stellar mass.
A caveat is that the Ridgway \et results may not reflect the properties of the
$2 - 3$ magnitude more luminous radio-quiet quasars observed at $z \approx 4$,
but deep imaging of $z > 3$ radio-quiet quasar hosts is currently unavailable.


\paragraph{Can AGNs Occur in Young Galaxies?}

Probably the most convincing evidence that relatively young galaxies 
at $z \gsim 3$ are capable of hosting massive, active black holes comes,
somewhat surprisingly, from the distant, small-scaled, young, star-forming 
Lyman-break galaxies (LBGs; \eg Steidel \et 1996, 1999).
The presence of an active black hole can be inferred for a tiny fraction 
($\sim$1\%) of the $\sim$1000 LBGs detected in a survey at $z \approx 3$ 
(Steidel \et 2002) by the presence of broad emission lines in their rest-UV 
spectra. Specifically, the average spectrum of these broad-lined LBGs
(Steidel \et 2002) suggests that the mass of the black hole is typically 
rather high ($\gsim 10^8$\Msol): 
Table~\ref{lbg_agn.tab} lists estimates
of the average values of \mbh{}, \lbol{}, and \lol{} 
for different cosmologies and assumed dust extinction factors.
Based on the uncorrected observed fluxes, the broad-lined LBGs have
masses \mbh{} $\approx 10^8$\Msol{}, lower 
than the higher luminosity quasars at any redshift beyond 1.5 
(Figure~\ref{m_zlbins.fig}), and they are about an order of magnitude 
less luminous (Figure~\ref{l_zmbins.fig}). 
In spite of the large uncertainties, the inferred average \mbh{} value 
is intriguingly large considering 
(a) the relatively faint nature of both the nuclear and host 
galaxy emission, 
(b) the star-formation activity in the host galaxy (Steidel \et 1996), 
and (c) the low stellar mass in the galaxies 
at $z \approx 3$: they typically only contain $\sim$10\% $-$ 30\% of their 
expected final stellar mass (Ridgway \et 2001; Papovich \et 2001).
This suggests that the massive central black hole in the LBGs formed 
early and fast compared to the underlying stellar component.

\setcounter{table}{0}

\begin{table*}[t]
\begin{center}
\begin{minipage}{12cm}
\caption[]{Estimated$^a$ mass, luminosity, and Eddington ratio of the average broad-lined
AGN among Lyman-break galaxies.
\label{lbg_agn.tab}}
\end{minipage}
\begin{tabular}{cccccc}
\\
\hline \hline
\multicolumn{1}{c}{ Extinction} &\mbox{$H_0$}&
{$\log L_{\nu}$(1350\AA)}&
{$\log M_{\rm BH}$}&
{$\log L_{\rm bol}$}&
{$L_{\rm bol}/L_{\rm Edd}$}\\
\multicolumn{1}{c}{Factor}&
{(km\,s$^{-1}$)}&
{(erg s$^{-1}$ Hz$^{-1}$)}&
{($M_{\odot}$)}&{(erg s$^{-1}$)}&{ }\\
\multicolumn{1}{c}{Applied}&{Mpc$^{-1}$)}&&&&\\
\multicolumn{1}{c}{(1)} &
\multicolumn{1}{c}{(2)} &
\multicolumn{1}{c}{(3)} &
\multicolumn{1}{c}{(4)} &
\multicolumn{1}{c}{(5)} &
\multicolumn{1}{c}{(6)} \\
\hline
None & 50 & 29.54$\pm$0.02 & 8.2$\pm$0.1 & 45.56$\pm$0.2 & 0.20$\pm$0.10 \\
None & 75 & 29.19$\pm$0.02 & 7.9$\pm$0.1 & 45.20$\pm$0.2 & 0.15$\pm$0.08 \\
50   & 50 & 31.24$\pm$0.02 & 9.4$\pm$0.1 & 47.26$\pm$0.2 & 0.64$\pm$0.32 \\
50   & 75 & 30.89$\pm$0.02 & 9.1$\pm$0.1 & 46.90$\pm$0.2 & 0.50$\pm$0.25 \\
\hline
\hline
\end{tabular}
\begin{minipage}{15cm}
\vspace{0.2cm}
{\small
$^a$ These estimates and measurement uncertainties are based on the composite 
spectrum of Steidel \et (2002).
While it is based on non-spectrophotometric data, it should after all be
representative of the mean brightness level of these AGNs (C.\ Steidel, 2002, 
private communication), sufficient for the discussion here.  Specifically, the 
typical \civ{} line width is measured (from the published plot and similarly
to the other data analyzed here; \S~\ref{data}) to be 
4670$\pm$500\,\kms, and the 1350\AA{} flux is (2.01$\pm0.09$) $\mu$Jy.  
Since the composite spectrum shows evidence of some dust reddening, the \mbh{},
\lbol{}, and \lol{} values were also computed after making a correction for
dust extinction. Steidel \et (2002) estimate an extinction factor of $\sim 50$
for their narrow-lined AGNs, but do not quote such a factor for the broad-lined
AGNs among their LBGs. Therefore, the estimate for the narrow-lined AGNs were
adopted here in attempts to bracket reality somewhat. 
In addition to the listed values of $H_0$, $q_0 = 0.5$ is assumed.
}
\end{minipage}
\end{center}
\end{table*}

A possible caveat is the uncertainty as to whether or not these faint LBG-type 
AGNs have similar SEDs to the bright quasars, on which both the \mbh{}
and, especially, \lbol{} estimates are based. But among the
lower-luminosity AGNs in the nearby Universe, only the faint, very
low-luminosity LINERs display dramatically different SEDs (Ho 1999),
mainly missing the ``big blue bump'' emission from the accretion disk. 
Seyfert galaxies, with which the LBGs with broad high-ionization 
lines may be most similar, display only minor SED differences in the 
optical-UV region compared to the more luminous quasars (\eg Sanders 
\et 1989; Elvis \et 1994). Moreover, 
the relatively large \civ{} line widths (4670$\pm$500\,\kms; very 
similar to the mean FWHM(\civ) of the luminous $z \approx 4$ quasars) 
and line equivalent widths (see Figure~1 by Steidel \et) 
support a high central mass (as argued in \S~\ref{zdistr}), in spite 
the faint nature of the galaxies.  Therefore, the derived mass and 
luminosity values are crude, but probably not too unrealistic.

\vskip 0.3cm \noindent
\subsubsection{Discussion and Summary}

Although largely circumstantial, the evidence presented above suggests 
that the $z \gsim 3$ massive, active black holes typically reside in 
relatively young (possibly dynamically unrelaxed) galaxies in their major 
star-forming phase.
The implication is thus that the black hole either formed long before
the stars in the galaxy or sometime during the first star-formation phase
and thereafter grew so fast relative to the stellar
mass that the nucleus shined long before the galaxy was fully formed.
Rix \et (2001) and Omont \et (2001) independently reached similar conclusions
based on different arguments. 
Rix \et find $z \approx 2$ radio-quiet quasars to have either 
higher nuclear luminosities or fainter host galaxies than lower-$z$ AGNs. 
They argue that the most plausible explanation is that the ratio of \mbh{} 
to the stellar mass increases with redshift.
Omont \et (2001) use millimeter data on $z \approx 4$ quasars to argue for 
a higher black-hole growth rate relative to the surrounding bulge stellar 
component in the host galaxy at early epochs, consistent with the current 
study and that of Rix et al. 
Such an evolutionary scenario is consistent with some hierarchical
models (\eg Haehnelt \& Rees 1993). Also, Silk \& Rees (1998) suggest that 
early black-hole growth might in fact help regulate star formation in the host 
galaxy and its evolution.
The overall conclusion reached here is thus that black holes likely
reach maturity before the stellar populations in their host galaxies.

Archibald \et (2002) claim that quasar activity is delayed 0.5 $-$ 0.7\,Gyr 
from the onset of star formation, during which $\gsim$75\% of the stars are 
formed, in apparent contradiction to the results presented above. 
However, Archibald \et assume that black holes grow by accretion with a 
radiative efficiency ($\eta$) of 10\%, while the value of $\eta$ and its
constancy with time (or redshift) are highly uncertain. A lower radiative
efficiency allows faster black-hole growth, and indeed, some 
recent estimates imply $\eta \approx 0.03$ (Schirber \& Bullock 2003).
Furthermore, one can equally well expect black-hole formation and early
mass growth to be very fast, as expected for the first generation of
massive stars (\eg Abel, Bryan, Norman 2002). 
Silk \& Rees (1998) find it unlikely that the first black 
holes formed with \mbh $< 10^6$ \Msol{}; in that case, it would take 
$\leq$0.2\,Gyr to generate a $10^8$ \Msol{} black hole from a $10^6$ 
\Msol{} seed accreting with $\eta \lsim$10\%, consistent with quasar
activity existing in very young galaxies, as it takes 0.3\,Gyr to form
50\% of the stars in the galaxy (Archibald \et 2002).

It is pertinent to re-emphasize that the presence of young starbursts in 
high-$z$ quasars cannot be inferred from their high dust content alone: 
a young evolving galaxy maintains a large dust mass ($\gsim 10^8$\Msol) 
from 0.04\,Gyr to 1.5\,Gyr after the initial starburst (\eg Archibald \et 2002).
But the combination of high masses of cool dust and cold (molecular) gas 
is much more convincing. 
Therefore, it is important for future investigations to study the cold gas 
contents in high-$z$ AGN host galaxies to test whether or not they are
in the process of building up the majority of their stars. Furthermore,
better coverage of the radio through near-IR SEDs of high-$z$ quasars
is required to constrain the relative contributions of AGN and starburst
heated dust for more accurate estimates of SFRs.

\vskip 0.3cm \noindent
\section{Conclusions \label{conclusion}}

The main results of this work can be summarized as follows:

(1)  
It is confirmed that the high-$z$ quasars have central masses of
at least a few times $10^9$\Msol{} as previously estimated from their 
luminosities (Fan \et 2001c). 
The masses, bolometric luminosities, and Eddington
ratios of the $z \gsim 3.6$ quasars are very similar to those of their 
lower-redshift cousins. Relative to the evolved \lbol{} and \mbh{} the 
$z \approx 4$ quasars are much more luminous and more massive than the typical 
quasars at $z < 3.5$. This assumes that the luminosity function, based on
the Large Bright Quasar Survey and the 2-Degree Field Quasar Redshift 
Survey, and its extension to $z \approx 4$ is reasonably reliable
(\S~\ref{zdistr}).

(2) 
The scaling laws from which the black-hole mass is estimated are fully 
applicable to distant luminous quasars. This is argued based on the 
similarity of quasar and AGN  spectral properties (line widths, line 
equivalent widths, line ratios, SEDs) at low and high redshift.
In particular, photoionization models confirm that if the 
radius\,$-$\,luminosity relationship does not extend to high-$z$ quasars, 
their spectra would look different than they do.
Mass estimates, based on scaling relationships, are not likely inaccurate 
by a factor greater than \mbox{$\sim$3 to $\sim$4} (\S~\ref{mreliability})
when applied to statistically significant samples.

(3)
There is a real ceiling to the \mbh{} and \lbol{} values at all redshifts:
\mbh{} $< 10^{10}$\Msol{} and \lbol{} $< 10^{48}$\ergs. These upper limits
are consistent with the recent studies of Netzer (2003) and Bechtold \et
(2003) to within the uncertainties and differences in adopted cosmologies.
Moreover, the large \mbh{} values at high-$z$ are not inconsistent with 
local \mbh{} values given the much smaller volume probed locally (\S~\ref{bhprops}).

(4) 
Black holes reach maturity (\ie reach masses \mbh{} $\approx 10^8 - 10^9$\Msol) before their
host galaxies: AGN host galaxies at $z \gsim$ 3 do not typically appear fully
assembled, dynamically relaxed, and/or the majority of their stars are not
formed or are somewhat evolved ($>$1 Gyr). 
Since the evidence for relatively young quasar-host galaxies at $z > 3$ is
mostly indirect, this interpretation needs to be tested further. For example,
complete coverage of the radio to IR SEDs of large representative samples
of $z > 3$ quasars will help provide improved constraints on the relative
starburst and AGN contributions to the far-IR emission. Detailed studies
confirming the correlation between large dust masses, large reservoirs of
cold molecular gas, and high star-formation rates in larger quasar samples
are also required. No doubt, the combination of {\it SIRTF} and the planned
{\it FIRST/Herschel} and ALMA telescopes will play an important part
in shedding light on these issues.

\acknowledgments

Thanks are due to Anca Constantin, Fred Hamann, Joe Shields, and Xiaohui Fan for 
providing the digital spectra of their $z \approx 4$ quasars, Martin Elvis for 
providing a digital form of the average quasar SEDs, Dave Sanders for providing 
his \lbol{} measurements of the Palomar-Green quasars, and Adam Steed for 
updating the radio-quiet SED.
Luis Ho, Kirk Korista, and Pat Osmer are all thanked for discussions 
or helpful comments.  Special thanks go to Brad Peterson for countless 
extensive and very valuable discussions, and for comments on the manuscript.
I am also thankful that Luis Ho, Pat Osmer, and Brad Peterson were brave enough 
to read and comment on an early, larger version of the manuscript; their input 
was very helpful and much appreciated.
The author gratefully acknowledges financial support from the Columbus Fellowship.
Part of the observations reported here were obtained at the MMT Observatory, a 
joint facility of the Smithsonian Institution and the University of Arizona.
This research has made use of the NASA/IPAC Extragalactic Database (NED)
which is operated by the Jet Propulsion Laboratory, California Institute
of Technology, under contract with the National Aeronautics and Space
Administration.


\end{document}